# Direct electrocaloric characterization of ceramic films

Uros Prah,[1,2†*] Matej Sadl,[2,3†] Alvar Torello,[1] Pierre Lheritier,[1] Veronika Kovacova,[1] Hana Ursic[2,3*] and Emmanuel Defay[1*]

[1] Materials Research and Technology Department, Luxembourg Institute of Science and Technology, Belvaux, Luxembourg.

[2] Electronic Ceramics Department, Jozef Stefan Institute, Ljubljana, Slovenia.

[3] Jozef Stefan International Postgraduate School, Ljubljana, Slovenia.

[†] The authors contributed equally to this work.

* uros.prah@list.lu, emmanuel.defay@list.lu, hana.ursic@ijs.si

## Abstract

Reliable and accurate characterization of the electrocaloric effect is necessary to understand the intrinsic properties of materials. To date, several methods have been developed to directly measure the electrocaloric effect. However, each of them has some limitations, making them less suitable for characterizing ceramic films, which rely almost exclusively on less accurate indirect methods. Here, a new approach is proposed to address the process of rapid heat dissipation in ceramic films and to detect the electrically induced temperature change before it thermally bonds with the surrounding elements. By using a polymer substrate that slows heat dissipation to the substrate and fast infrared imaging, a substantial part of the adiabatic electrocaloric effect in $Pb(Mg_{1/3}Nb_{2/3})O_3$-based ceramic films is captured. Infrared imaging provides a robust technique to reduce the ratio between the adiabatic and the measured electrocaloric temperature change in micrometer-sized ceramic films to a single-digit number, ~3.5. The obtained results are validated with another direct thermometric method and compared with the results obtained with an indirect approach. Despite different measurement principles, the results obtained with the two direct methods agree well. The proposed approach is timely and could open a door to verify the predicted giant electrocaloric effects in ceramic films, thus accelerating the process of their integration into functional devices.

## Introduction

About a century after its discovery and the first experimental confirmations, solid-state cooling based on caloric effect has emerged as one of the most promising alternatives to the less energy-efficient and ecologically problematic vapor-compression cooling technology. The caloric effect refers to the reversible adiabatic temperature ($\Delta T_{ad}$) and isothermal entropy ($\Delta S_{iso}$) changes of the material upon application of external stimuli, such as electric field (electrocaloric), magnetic field (magnetocaloric), or stress (mechanocaloric effect).[1–3] Of all the external stimuli, the electric field is relatively easy and cheap to generate, so the electrocaloric (EC) effect is of particular interest for miniaturization and mass production.[4] In addition, external energy recovery can be integrated into the EC circuit to further increase the energy efficiency of the EC cooling device.[5] However, the bottle neck towards further commercialization of EC materials is generally the low dielectric strength of EC ceramics, which prevents the application of higher electric fields and thus the achievement of large EC effects.[6,7] One way to overcome this is to fabricate thin ceramic films, which, due to their typically very uniform and high-quality structure, lead to significantly improved dielectric strength. This allows the application of much higher electric fields, yielding to so-called giant EC effects (>10 K). Because of their small size, EC films can be driven hard at a relatively low voltage, making them good EC active bodies for downsized cooling applications, such as cooling flexible and portable electronics.[8-10]

Interest in the EC effect and materials was revived with the discovery of giant EC effects in ceramic thin films when Mischenko et al. demonstrated in 2006 EC temperature changes ($\Delta T_{EC}$) of 12 K in sub-micrometer thick $Pb(Zr_{0.95}Ti_{0.05})O_3$ films.[11] Later, many more studies on ceramic and polymer thin films followed, where the EC effect was determined almost exclusively indirectly from polarization data using the well-known Maxwell relation.[12-17] Due to its simplicity and ease of application, the so-called indirect Maxwell method has become widely accepted in practice.[6,18,19] However, this approach relies on numerous assumptions that are prone to error, which is reflected in the typically large discrepancies between reported studies that always cast doubt on the results obtained. Despite all this, indirect methods, when properly applied, can be successfully used to estimate the magnitude of the EC effect, but they always require further validation by the direct EC measurement methods.

Direct EC measurement methods refer to the precise measurement of temperature or isothermal heat ($Q$) induced by an electric field, typically performed by thermometry or calorimetry, respectively.[6,18,20] To date, few studies have addressed the direct EC characterization of ceramic thin films and structures. Various techniques have been used, such as high-resolution calorimetry with a small bead thermistor,[20-23] a heat flux sensor,[24] a thin-film resistance thermometer (3$\omega$),[25,26] scanning thermal microscopy,[27] and laser-based approaches[28] (for more details, see SI, Table S1). Each of the above approaches has some drawbacks, such as long response time, many post-processing steps, expensive equipment, or limitations in the application of the desired electric field signals, which makes them less universal and more difficult to perform. Therefore, despite these promising attempts,[21,24-28] the EC community is still looking for a simple, reliable, and convincing direct EC measurement method to undoubtably confirm the giant EC effects in thin films predicted by indirect methods.

Yet, direct determination of $\Delta T_{EC}$ in thin films deposited on a substrate is very difficult. Due to the low thermal mass of the EC thin film, the EC induced heat (or $\Delta T$) immediately dissipates in the surrounding elements, such as the electrodes and the substrate, which acts as a large heat sink/source.[6,18,19] Due to the rapid heat exchange between the active and inactive parts of the sample stack, it is very difficult (or impossible) to achieve the adiabatic conditions and thus capture the intrinsic EC effect. For example, an intrinsic $\Delta T_{EC}$ of 10 K may dissipate completely or drop to a few mK within a millisecond. Consequently, large correction factors (i.e., ratios between $\Delta T_{ad}$ and measured $\Delta T_{meas.}$) are required, which increases measurement uncertainty and thus confidence in the measurement method. In general, the discrepancy between the measured and the intrinsic (adiabatic) $\Delta T_{EC}$, and thus the magnitude of the correction factor ($k$), is directly related to the difference in thermal masses between EC active material and the surrounding EC inactive components, in our case mainly the substrate. From previous reports (Fig. 1), $k$ in thick films (>10 μm) can reach values of several tens to several hundreds, while the magnitude of $k$ in submicron thick films can increase to several thousands. Note that because of the lower thermal conductivity, as well as the ability to make freestanding layers, direct EC measurements of polymer films are less problematic and generally do not require a large $k$ (see Fig. 1 and SI, Table S1 for comparison).[29-32]

There are two ways to address the problem of rapid heat dissipation in ceramic thin films: (i) slowing down the heat dissipation process by changing the properties of the selected substrate material (i.e., thermal mass and thermal diffusivity) and (ii) using a fast thermometry technique to capture a larger fraction of the EC induced $\Delta T$ before it is fully dissipated into the substrate. To satisfy the above two conditions, we have developed an approach to capture a significant portion of the intrinsic EC effect that involves a polymer substrate with low thermal conductivity and the use of fast infrared (IR) imaging (both contributions are marked with arrows in Fig. 1). To keep the EC induced heat in the film longer, we replaced the metal substrate (e.g., the commonly used Si wafer) with polyimide, whose thermal conductivity is one order of magnitude lower than that of ceramics and three orders of magnitude lower than that of metals (SI, Fig. S7). However, the replacement of the substrate material is still not sufficient to capture the substantial part of the EC effect in ceramic films (see the upper blue arrow and the red dot in Fig. 1). Therefore, a fast IR camera was used, which allows simple non-contact temporal and spatial temperature measurements at high frequency (kHz range).[6,18,19,33] Moreover, in this case, no additional sample preparation steps are usually required, making the IR camera a compact and universal measurement method that can be used for direct EC characterization of ceramic films (as demonstrated here) as well as for testing the performance of cooling devices.[34,35]

Thanks to the room-temperature (RT) aerosol deposition (AD) method, we were able to prepare dense and flexible ~3 μm thick $0.9Pb(Mg_{1/3}Nb_{2/3})O_3$–$0.1PbTiO_3$ (PMN–10PT) ceramic films on 125 μm thick polyimide substrate. In these films, we were able to capture almost one-third of the intrinsic EC effect using a commercially available IR camera and reduce correction factor to a single-digit number, namely 3.1–3.5 (see the bottom blue arrow and the red dot in Fig. 1 and SI, Table S1). The intrinsic EC effect was finally derived by a 2D numerical heat transfer model using finite element modeling (FEM). The combination of fast IR imaging and numerical modeling in our PMN–10PT ceramic films deposited on a polyimide substrate resulted in $\Delta T_{EC}$ ~1.65 K at RT (25 °C), while the maximum (~2.45 K) was reached at 99 °C (both at 180 V or 600 kV cm$^{-1}$). The results obtained were compared and validated with another commonly used direct thermometric method, namely, a modified calorimeter with a small thermistor.[6,20] Despite the completely different measurement principles, i.e., tracking thermal relaxation on different time scales, the results obtained with the two direct methods agree well for the prepared film structure. The advantages and disadvantages of both methods are discussed, highlighting their limitations and predictions for application in other thin film systems.

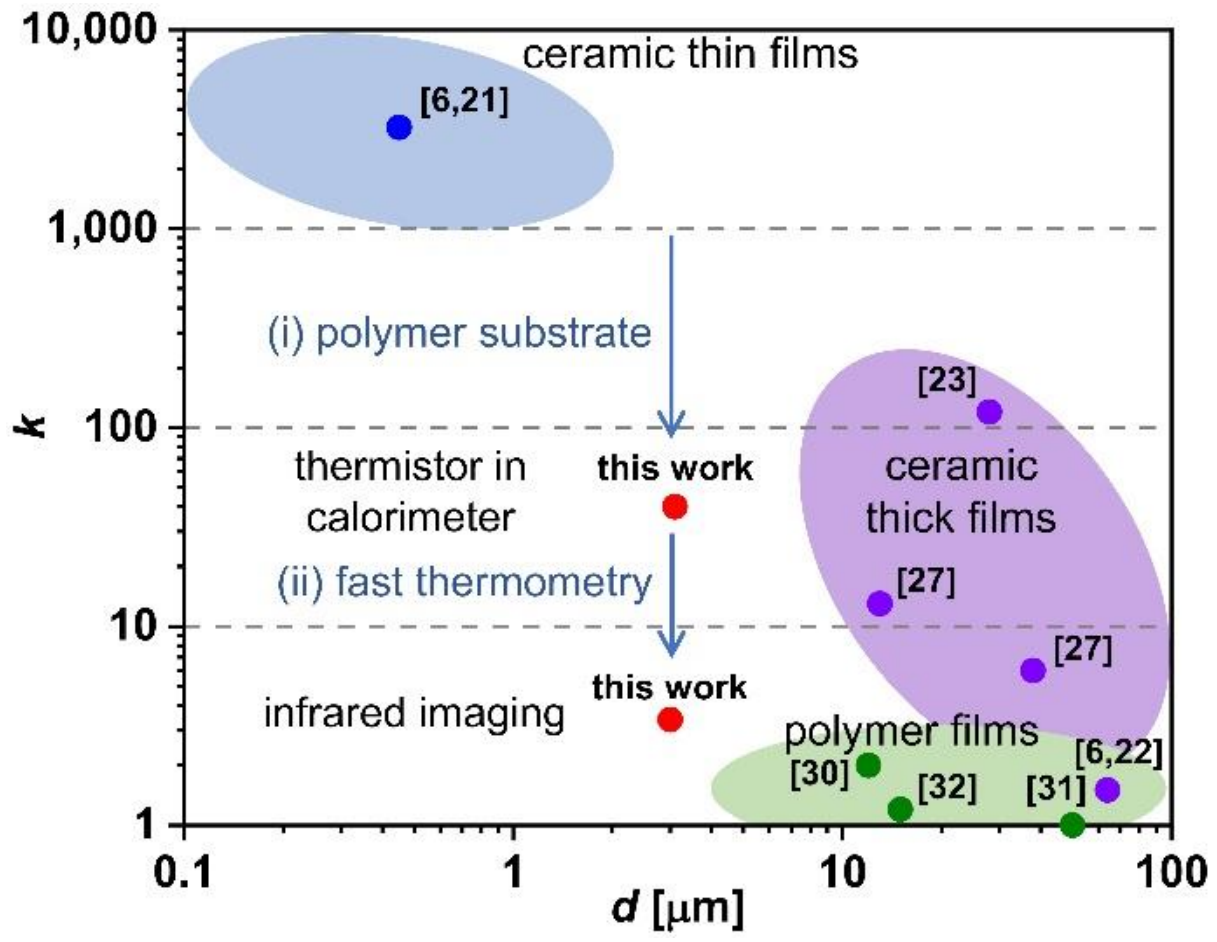

Figure 1. Correction factor ($k$) as a function of film thickness ($d$) for various ceramic and polymer films, derived from reported studies. The results obtained in this work are marked with red dots, while the blue arrows indicate the path used to obtain these results.

# Results

## Sample characteristics

The sample stacks used in this study consist of a flexible polyimide substrate, PMN–10PT ceramic film, and bottom and top Au electrodes (Fig. 2a). For direct EC measurements, it is important that the active (electrode-covered) surface area is large enough to maximize the ratio of EC active to inactive thermal masses (see Methods section for more details). Structural and microstructural characterization of the sample stacks shows that the PMN–10PT ceramic films are of high quality, making the room-temperature aerosol deposition method a powerful tool for the fabrication of ceramic films on various low melting point substrates and flexible materials. X-ray diffraction (XRD) and scanning electron microscopy (SEM) reveal a single-phase (i.e., pyrochlore-free) material with uniform, homogeneous and dense microstructure (details in SI, Fig. S1).

The fabricated PMN–10PT ceramic films exhibit relaxor-like ferroelectric behavior, as shown by the diffuse temperature-dependent permittivity data with a weak frequency dispersion and slim, tilted, and unsaturated polarization–electric field ($P$–$E$) hysteresis loops (SI, Fig. S2). Relaxor ferroelectrics, such as PMN-rich PMN–PT ceramics, have been shown to be advantageous for EC cooling applications due to the wide temperature range with large EC effects, large change in $P$ as a function of $E$ and $T$, and low EC fatigue.[36-38] Despite the relatively large electrode surface area, the samples withstand an applied electric field of up to 700 kV cm$^{-1}$ and achieve a maximum polarization ($P_{max}$) of ~34 µC cm$^{-2}$ at RT (SI, Fig. S2b). The electrical properties obtained are comparable to the results obtained with PMN–10PT ceramic films deposited on stainless steel,[39] proving that high quality samples can also be fabricated on flexible polymer substrates and thus can be used for further direct EC characterization.

## Determination of electrocalorically induced $\Delta T$ with infrared imaging

Direct measurements of the EC effect were performed using fast IR imaging. Due to the high thermal reflectivity of the Au electrode, an additional black ink layer is mandatory to increase the emissivity, allowing for accurate monitoring of the EC induced $\Delta T$ (see Methods section and SI, S4 for more details). The entire sample stack used is shown in Fig. 2a. Typical EC cycles obtained from the selected black ink-coated areas marked with crosses in the IR thermogram (SI, Fig. S5a), are shown in Fig. 2b. The EC cycle consists of adiabatic heating and cooling peaks when an external electric field is applied and removed quickly, respectively. The high IR camera frame rate of 1.35 kHz was used to detect fast changes of temperature, resulting in a temporal resolution of ~0.74 ms, while EC induced $\Delta T$ heating and cooling peaks were obtained at ~1.5 ms after electric field application, corresponding to adiabatic conditions (SI, Fig. S4). To obtain reliable and repeatable results, a series of four consecutive EC cycles were measured for each selected ambient temperature and electric field strength. As can be seen in Fig. 2b, a slight asymmetry between the intensities of the EC heating and cooling peaks occurs at higher fields and temperatures, which could be related to self-heating of the material due to increased hysteresis losses.[36,38]

The intensity of EC cooling peaks measured at different ambient temperatures and applied electric fields are shown as the mean and standard deviation of four EC cycles in Fig. 2c. The temperature-dependent measurements of the cooling effect showed an increasing trend with increasing temperature, with $\Delta T$ of 0.50 K at 22.4 °C and a peak of 0.72 K at 99 °C (both values were obtained at 600 kV cm$^{-1}$). Note that the slight scatter in the measured $\Delta T$ values is a consequence of the measurement setup and the associated stabilization of the ambient temperature, as well as the fact that we follow the fast dynamics of heat transfer between the active film and the substrate (see Methods section and SI, S4 for more details). The monotonic increase in $\Delta T$ and thus the absence of intermediate peaks over the entire measured temperature range (15–100 °C) is consistent with the broad and diffuse peak-permittivity data (SI, Fig. S2a) and with previous studies of thick film,[27] multilayer capacitor,[40] and bulk[37] PMN–10PT ceramics.

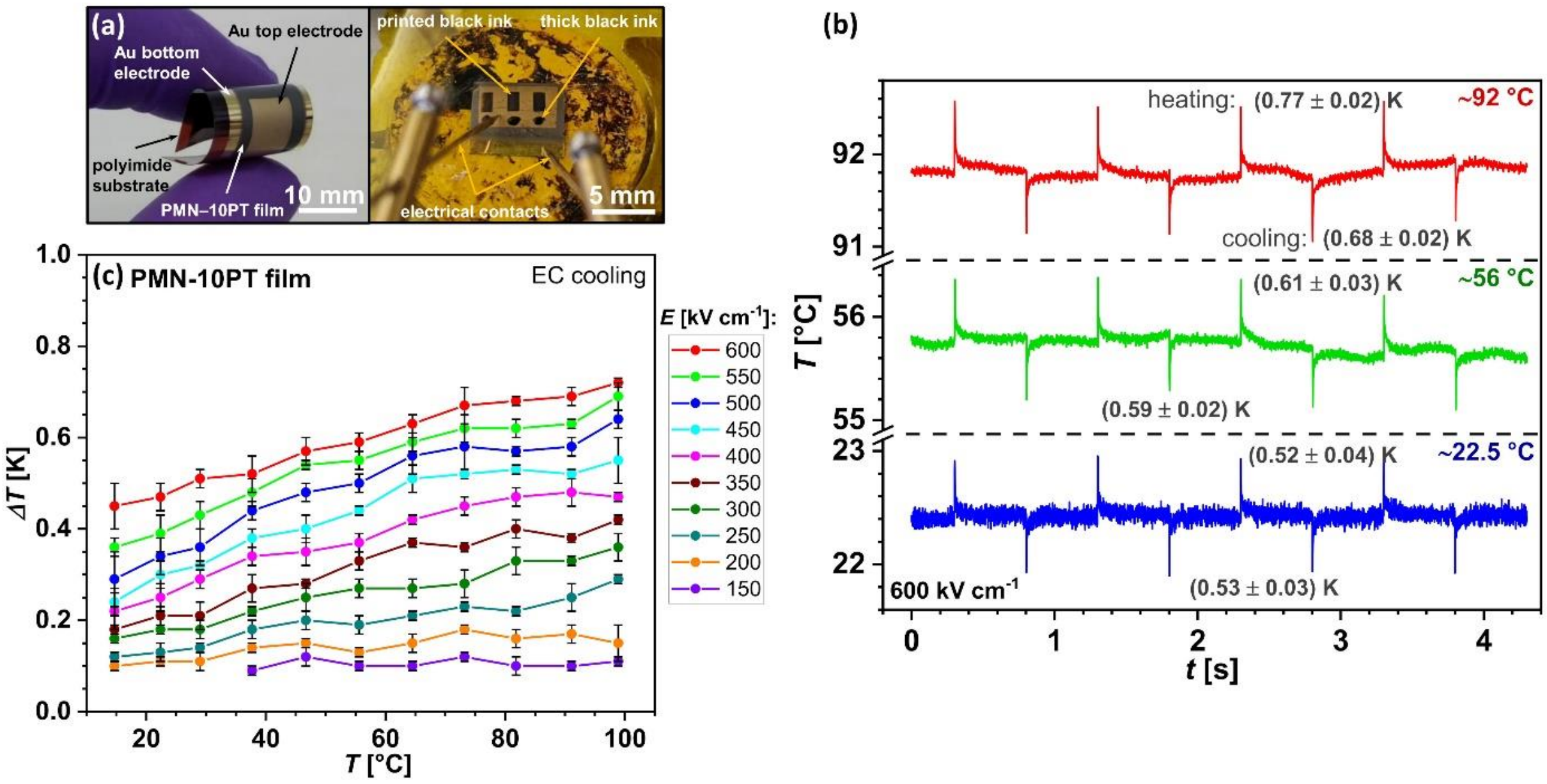

Figure 2. (a) Photographs of the PMN–10PT sample stack used for IR imaging, (b) sequence of four EC cycles measured with the infrared camera at different ambient temperatures and 600 kV $cm^{-1}$. (c) Temperature- and field-dependent $\Delta T$ corresponding to the measured cooling peaks originating from the printed black ink layers.

## Heat flow analysis and determination of intrinsic (adiabatic) $\Delta T_{EC}$

To gain insight into the dynamics of heat dissipation and thus predict how close we are to adiabatic conditions, a heat flow analysis was performed using a finite element method (FEM). Due to the rapid dissipation of EC induced heat mostly in the substrate, which occurs on a time scale smaller than the resolution of the IR camera (SI, Fig. S4), the $\Delta T$ values read on the printed black ink layer are underestimated. To evaluate the intrinsic $\Delta T_{EC}$ occurring in the active layer of PMN–10PT, we performed 2D finite element modelling of heat dissipation. The EC effect was modelled as a heat pulse acting on the active layer followed by heat dissipation to the surrounding components. In addition, the model was refined by a subsequent fit to the experimental data shown in Fig. 3a (see Methods section and SI, S5 for more details).

The 2D temperature profile model of a part of the sample stack after triggering the EC heat pulse, from which the temporal temperature change in different components of the sample stack was obtained, is shown in the inset of Fig. 3a. Fitting the numerical model to experimental data measured at three different temperatures allows us to reproduce the experimental curves and track the intrinsic EC effect in the ceramic layer alone. For simplicity, only positive heat pulses were refined. As shown in Fig. 3a, the correction factor was obtained by comparing the ratios between the numerically obtained $\Delta T_{ad.}$ inside the film and the experimentally measured $\Delta T_{meas.}$ on the top of the black ink layer. The correction factor obtained from three different temperatures over the whole measurement range is between 3.1 at RT and 3.5 at ~92 °C (Fig. 3a), which means that almost one-third of the intrinsic EC effect could be captured by the proposed measurement approach. The temperature dependence of correction factor over the entire EC measurement temperature range, combining modelled and interpolated data, is shown in SI, Fig. S9. Finally, by combining the results of the IR camera measurements (Fig. 2c) with a correction factor derived from modeling (SI, Fig. S9), we can estimate the intrinsic $\Delta T_{EC}$ in the PMN–10PT ceramic film, resulting in a $\Delta T_{EC}$ of ~1.65 K at RT, while the maximum of ~2.45 K is reached at 99 °C and 600 kV cm$^{-1}$ (Fig. 3b).

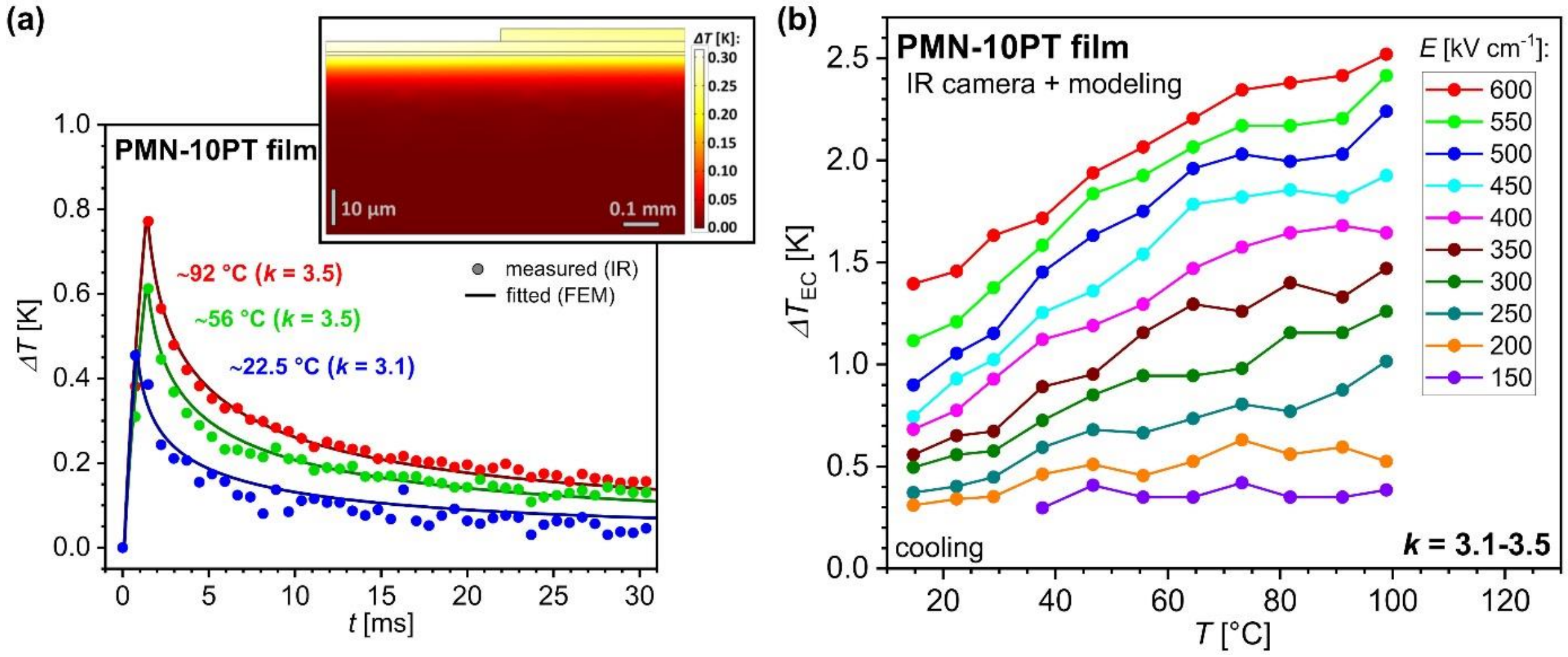

Figure 3. (a) Results of finite element modeling of the EC induced heat and its dissipation within the sample stack, fitted to the experimental data from which the intrinsic $\Delta T_{EC}$ values and corresponding correction factor ($k$) were derived. The inset shows the 2D temperature profile model of the sample stack after triggering the EC heat pulse. (b) Temperature- and field-dependent $\Delta T_{EC}$ values determined by IR camera measurements and numerical modeling.

## Discussion and Conclusions

The obtained $\Delta T_{EC}$ results were verified with another direct thermometric EC measurement method, called the thermistor-in-calorimeter method, which is often used to characterize bulk ceramics,[37,41] multilayer capacitors[40,42] as well as ceramic films[21,22] (for more details on this method, see SI, S6). The two direct methods compared are based on different measurement principles. While the thermistor-in-calorimeter method is "blind" during the first 0.5-1 s, when the internal thermal relaxation of the EC induced heat takes place, the IR camera focuses exactly on this time scale and tracks the rapid heat exchange between the EC active film and the substrate (Fig. 4a). Despite the much longer response time of the thermistor compared to a fast IR camera (about one second compared to one millisecond) and consequently an order of magnitude larger correction factor $k$ (i.e., ~40 compared to ~3.5), the obtained results show quite good agreement (see comparison in Fig. 4b). Regardless of the difference in measurement uncertainty, both direct measurements yield a similar trend and magnitude of EC effects.

Moreover, the longer response time of the thermistor allows for the decoupling of the two contributions introduced in this work to solve the problem of fast heat dissipation in ceramic films, i.e., (i) the use of a substrate with low thermal conductivity and (ii) the fast thermometry measurement (indicated by blue arrows in Fig. 1). The comparison of the correction factors of the two used direct methods (see the two red dots in Fig. 1) shows that the use of a substrate with low thermal conductivity alone (upper red dot in Fig. 1) is not sufficient to bring the correction factor into single digits. On the other hand, using fast thermometry alone is also not enough to bring the correction factor below 10, which shows the importance of adapting the substrate material as well. This was confirmed by finite element modeling, where a hypothetical replacement of the polyimide substrate with $Al_2O_3$ ceramics (SI, Fig. S7) leads to an increase in the correction factor to ~30 (not shown here). These results highlight the importance of both contributions, i.e., (i) the polymer substrate and (ii) fast IR imaging, in capturing nearly one-third of the intrinsic EC effect in micron-scale ceramic films.

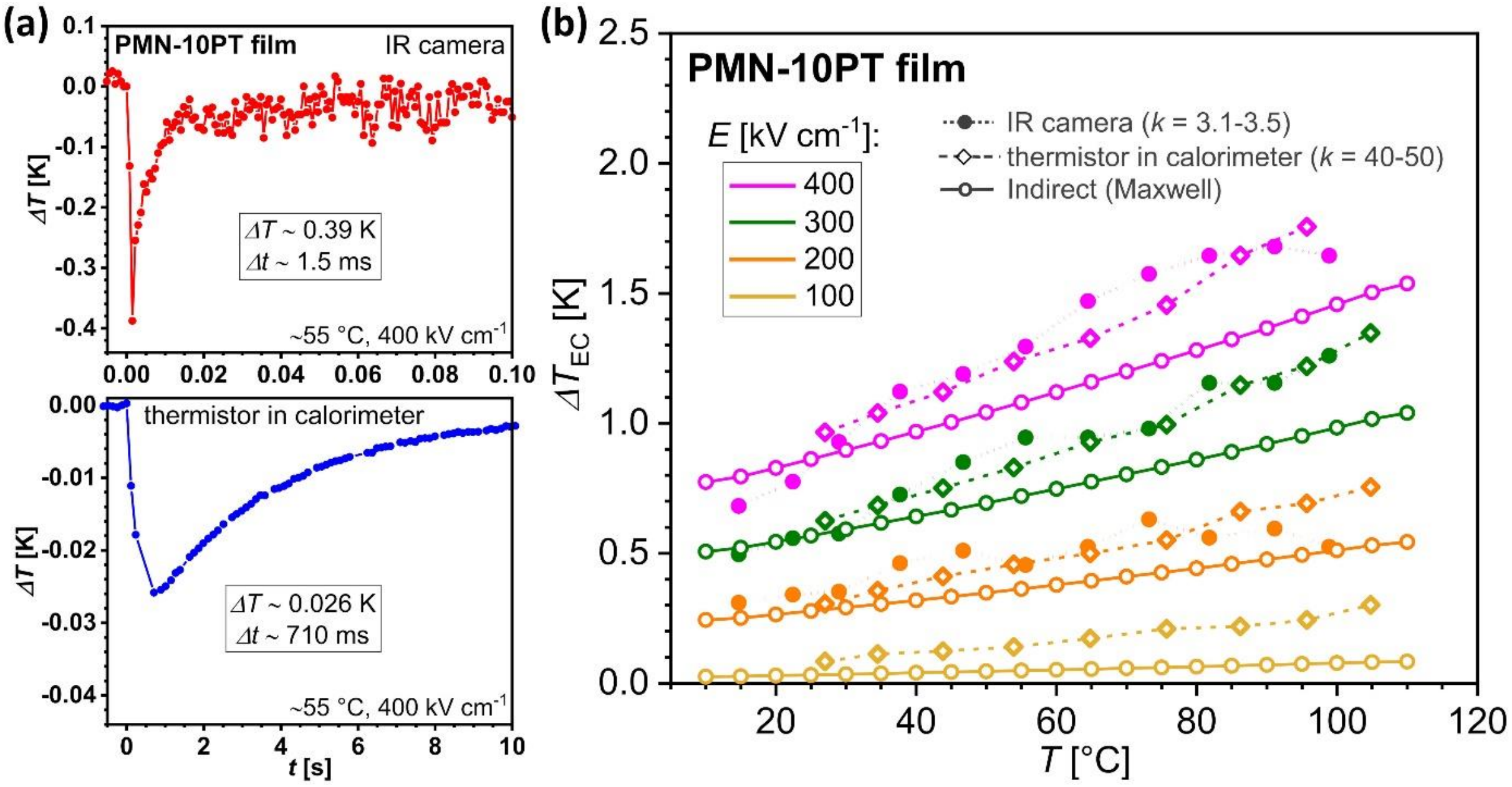

Figure 4. (a) Electrocaloric cooling peak measured with the infrared camera and the thermistor-in-calorimeter method at ~55 °C and 400 kV cm$^{-1}$. Note the difference of one order of magnitude in the temperature and time scales. (b) Comparison of $\Delta T_{EC}$ values obtained using two different direct (i.e., IR camera with numerical modeling and thermistor-in-calorimeter with correction) and indirect (exploiting Maxwell relations) measurement methods.

The advantage of the thermistor-in-calorimeter method is the good thermal stabilization and the precise determination of the EC induced $\Delta T$ (in the mK range). However, due to the long response time of the thermistor, only the external thermalization of the entire sample stack can be followed (Fig. 4a), which limits the application of the method to samples where the EC induced heat is still sufficiently large after thermalization of the thermistor (SI, S6). In addition to the larger correction factor compared to the IR camera measurements, its determination requires precise and time-consuming measurements of the size and mass of all the subcomponents involved in the EC measurement, which increases the measurement uncertainty of the method. On the other hand, the IR camera focuses on internal thermalization and enables us to follow the fast heat dynamics between the EC active film and the substrate (Fig. 4a). Consequently, the results in the measured $\Delta T$ are somewhat more scattered (i.e., larger noise), which can be partially addressed by averaging the results of multiple measurements. Here we are close to the limit of the thermistor-in-calorimeter method that makes it difficult to characterize thinner (<1 μm) ceramic films and structures, which means that the EC induced $\Delta T$ of the sample stack can be either negligible or very small, requiring correction factors of several 1000 (leading to a large error).[6,21] In contrast, IR camera measurements focusing only on the active film can still be extended to further reduce the size of the samples studied.

In addition, the results of the two direct EC methods were compared with the indirect Maxwell method (details in SI, S7). Although Maxwell's equations require ergodic system,[18,20,43] the indirectly obtained EC values of the PMN–10PT relaxor composition show relatively good agreement with both direct measurement methods (Fig. 4). More precisely, the results of the indirect method are slightly underestimated compared to the results of the two direct measurement methods. This confirms that indirect methods, when properly applied, can be used to estimate the magnitude of the EC effect simply by comparing the measured *P–E* loops, as shown in other studies on bulk[37] and multilayer[40] PMN–10PT ceramics. However, it should be noted that in this case the *P–E* loops were measured at a high frequency of 1 kHz to obtain comparable values, while the results at lower frequencies, namely 100 Hz (see SI, Fig. S16), are further from the values obtained directly. This once again underlines the importance of validating indirectly obtained results with direct methods.

In conclusion, we have shown in this work that changing the thermal properties of the substrate and using fast thermometry is an effective way to directly characterize the EC effect in ceramic films. The proposed direct measurement approach provides the opportunity for further improvements to characterize even thinner ceramic films (ideally <1 μm) by further reducing the thickness of the substrate material and reducing the measurement noise by improving thermal isolation and implementing lock-in operation during IR camera measurements. All these additional modifications open the possibility of using this method to characterize thin ceramic films and investigate directly the predicted giant EC effect demonstrated by indirect methods. This could be achieved, for example, by transferring submicron-thick ceramic films, which are usually fabricated on metal substrates, to materials with lower thermal conductivity, such as polymers, using lift-off techniques.[44] Therefore, the proposed measurement approach could represent a standard method for the future direct EC characterization of ceramic films.

## Methods

### Electrocaloric material preparation

PMN–10PT ceramic powder was prepared from oxides using mechanochemical-assisted synthesis. The pre-synthesized PT seeds and PMN powder mixture[45] were homogenized in stoichiometric molar ratio by wet ball milling (200 $min^{-1}$, 2 h), followed by high-energy dry milling in tungsten carbide milling vial filled with tungsten carbide milling balls (300 $min^{-1}$, 36 h). The same planetary ball mill (PM400, Retsch, Germany) was used for both processes. In order to tailor the morphology and particle size distribution required for successful aerosol deposition (AD), a series of additional post-synthesis treatments such as annealing (900 °C, 1 h) and milling were performed. The whole process of powder preparation and all post treatments are described in Ref. 39.

### Electrocaloric film fabrication

PMN–10PT films were prepared by the AD method at RT (InVerTec, Germany). Before deposition, the ceramic powder was sieved and dried (100 °C, 12 h in vacuum). During the AD process, the powder particles were mixed with a nitrogen carrier gas and formed an aerosol, which was subsequently ejected by a pressure difference and deposited on a substrate (deposition details in Ref. 46). A 125 μm thick polyimide foil (Kapton® HN, DuPont, USA) with a pre-sputtered ~1 μm thick Au bottom electrode (RF-magnetron sputtering equipment, 5Pascal, Italy) was used as substrate. The deposited films were post-annealed at 400 °C for 1 h to release the internal stresses generated during the deposition process and slightly increase the crystallite size. Finally, the top Au electrode was sputtered with a thickness of ~0.1 μm and an average surface area of ~20 $mm^2$. A photograph of the entire sample stack is shown in Fig. 2a. Further details on the preparation of the films can be found elsewhere.[39,47]

### Direct EC measurements using fast IR camera

Direct measurement of EC temperature change ($\Delta T$) was performed using a commercially available IR camera (FLIR X6580sc) with a quantum InSb detector and forced cooling. The IR camera was pre-calibrated by the manufacturer. Due to the high thermal reflectivity of the top Au electrode, additional black ink was used to bring the surface emissivity close to 1 and allow accurate monitoring of the sample temperature (SI, Fig. S5a). To minimize the amount of EC inactive materials, ~3.5 μm thick black ink layers were printed on the top Au electrode (see the whole sample stack in Fig. 2a), whose emissivity was calculated and corrected according to the thick black ink layer (black matt, Colorjelt RAL 9005, Jelt) assuming an emissivity of 1 (SI, Fig. S5b). The black ink layers were printed using a commercial office printer (Konica Minolta, bizhub C454e). The thickness of the black ink layers was determined using profilometry and optical microscopy (SI, Fig. S6). Note that according to the manufacturer's specifications, the emissivity must be higher than 0.5 to obtain reliable temperature data.[48]

To trigger the EC cycle, a square wave signal was applied with a function generator (Keithley 3390) and a high voltage amplifier (Trek 2220) and monitored with an oscilloscope (Agilent DSO5014A). Ambient temperature was controlled with a Linkam (THM S600) temperature stage with active cooling by liquid nitrogen (LNP95). A Python script was developed for fast, automatic operation and data acquisition. A schematic representation of the entire measurement setup can be found in SI, Fig. S3.

A high IR camera frame rate of 1.35 kHz and an integration time of 400 μs were used to ensure adiabatic conditions as much as possible. The measured EC temperature change was averaged over the selected ink-coated areas (marked with squares in Fig. 2a and crosses in SI, Fig. S5a) and plotted as the mean and standard deviation of four consecutive EC cooling peaks (Fig. 2b). Due to the "open air" environment, i.e., the sample placed directly on the hot plate-like temperature stage, the noise of the measured temperature was slightly increased (±0.1 °C).

**Numerical modeling of heat dissipation**

A heat flow analysis was performed by the finite element method (FEM) using the heat transfer module of COMSOL Multiphysics software (version 5.6). A 2D model was created assuming the dimensions of a real sample (see Fig. 3a). The modelled sample stack consists, in order from top to bottom, of the printed black ink layer, the PMN–10PT ceramic film, the bottom Au electrode, and the polyimide substrate. Due to its smaller thickness, the contribution of the top electrode is negligible and is therefore not included in the model for simplicity. The thicknesses of all components of a sample stack along with their properties (i.e., density, specific heat and thermal conductivity) used to build a numerical model are summarized in SI, Fig. S7. For a given sample stack, a user-controlled mesh with an element size of $1.2 \cdot 10^{-6}$ to $212 \cdot 10^{-6}$ with a maximum element growth rate of 1.1, a curvature factor of 0.3, and a resolution of narrow regions of 2 was chosen for the calculation (SI, Fig. S8).

The EC effect was modelled with a positive heat pulse applied to the active layer. The heat power density ($P_{EC}$) was derived from:

$$P_{EC} = \frac{\rho \cdot c_p \cdot \Delta T_{ad.}}{t_{pulse}}$$

where $\rho$ is density, $c_p$ is specific heat, $\Delta T_{ad.}$ is the adiabatic EC effect, and $t_{pulse}$ is the duration of a heat pulse. A time-dependent study based on heat transfer in the solid model was performed, taking into account the adiabatic boundary conditions on the external walls (i.e., negligible heat transfer to the air at a given time). The dissipation was modelled for a time range from 0 to 0.5 s. The obtained numerical model was further refined and verified by fitting the experimental data shown in Fig. 2b. This allowed us to reproduce the experimental curves and track the intrinsic EC effect in the ceramic films alone. The correction factor $k$, i.e., the ratio between the obtained $\Delta T_{ad.}$ inside the ceramic film and $\Delta T_{meas.}$ on the top of the black ink layer, was determined by FEM for three different temperatures over the whole range of measured temperatures (Fig. 3a). The modelled heat pulses always start at 10 μs, while they last 0.7 ms for the measurement at ~22.5 °C and 1.3 ms for the measurements at ~56 °C, ~92 °C. The temperature dependence of the correction factor was determined using the numerical model for three different temperatures (i.e., ~22.5 °C, ~56 °C and ~92 °C), while the correction factor was interpolated for intermediate temperatures (see SI, Fig. S9 for more details).

## Acknowledgements

U.P., V.K. and E.D. acknowledge the Fonds National de la Recherche (FNR) of Luxembourg for support of this work under the BRIDGES2020/MS/15410586/CALPOL/Defay project. H.U. and M.S. thank the Slovenian Research Agency (N2-0212, J2-3058, P2-0105, young researcher project PR-08977) and the JSI Director's Fund 2017-ULTRACOOL.

## Author contributions

H.U. and E.D. proposed the whole experimental study. M.S. prepared the samples and performed the structural and microstructural analysis and dielectric measurements. U.P. and P.L. performed the infrared imaging. A.T., P.L. and V.K. performed the finite element modeling. U.P. and V.K. measured ferroelectric properties and determined electrocaloric effect using the indirect method. M.S. performed thermistor-in-calorimeter measurements and analyzed the data with U.P. U.P. analyzed the data and wrote the manuscript with the assistance of all co-authors. H.U. and E.D. obtained funding and supervised the project.

## Conflict of interest

The authors declare no conflict of interest.

# Supplementary Information

## Direct electrocaloric characterization of ceramic films

Uros Prah,[1,2†*] Matej Sadl,[2,3†] Alvar Torello,[1] Pierre Lheritier,[1] Veronika Kovacova,[1] Hana Ursic[2,3*] and Emmanuel Defay[1*]

[1] Materials Research and Technology Department, Luxembourg Institute of Science and Technology, Belvaux, Luxembourg.
[2] Electronic Ceramics Department, Jozef Stefan Institute, Ljubljana, Slovenia.
[3] Jozef Stefan International Postgraduate School, Ljubljana, Slovenia.
[†] The authors contributed equally to this work.
* uros.prah@list.lu, emmanuel.defay@list.lu, hana.ursic@ijs.si

## Contents:

## S1.) Direct electrocaloric measurements of films – comparison with the literature

Table S1: Results of direct EC measurements on thin (<5 µm) ceramic films (highlighted in blue), thick (>10 µm) ceramic films (highlighted in purple) and polymer films (highlighted in green) from the literature and their comparison with the results of this study (highlighted in red).

| No. | Material | *d* [µm] | Substrate | Direct EC method | $\Delta T_{meas.}$ [°C] | $\Delta T_{EC}$ [°C] | *E* [kV cm$^{-1}$] | *k* | Ref. |
|---|---|---|---|---|---|---|---|---|---|
| 1 | PLZT [a] | 0.45 | 600 µm Si | thermistor in calorimeter | ~0.01 | ~40 | 1250 | 3230 | [1, 2] |
| 2 | PZT [b] | 0.15 | (Pr,Nd,Sm,Gd,Tb,Dy)$ScO_3$ | laser-based | * | 0.10 | 67 | * | [3] |
| 3 | | 0.12 | | 3$\omega$ | ~0.01 | * | 50 | * | [4, 5] |
| 4 | BST [c] | 3 | FTO (F:$SnO_2$) glass | heat flux sensor | * | 10.1 | 600 | * | [6] |
| 5 | PMN–10PT | 3.0 | 125 µm polyimide | infrared imaging | 0.72 | 2.45 | 600 | 3.4 | this work |
| 6 | | 3.1 | | thermistor in calorimeter | 0.037 | 1.46 | 380 | 40 | this work |
| 7 | PMN–10PT | 13 | PMN–10PT | scanning thermal microscopy | 0.23 | 1.8 | 308 | 13 | [7] |
| 8 | | 38 | | | 0.23 | 1.3 | 105 | 6 | |
| 9 | PMN–35PT | 64 | 10 µm Pt | thermistor in calorimeter | * | 1.6 | 80 | 1.5 | [2, 8] |
| 10 | PLZT [a] | 28 | $Al_2O_3$ | thermistor in calorimeter | * | 1.8 | 68 | >100 | [9] |
| 11 | P(VDF-TrFE) [d] | 0.4–6 | / | thermistor in calorimeter | * | 2.25 | 490 | * | [10] |
| 12 | P(VDF-TrFE-CFE) [e] | 11–12 | / | infrared imaging | 2.6 | 5.2 | 900 | 2 | [11] |
| 13 | P(VDF-TrFE-CFE) | 50 | / | infrared imaging | 2.75 | 2.75 | 800 | / | [12] |
| 13 | P(VDF-TrFE-CFE) [f] | 15 | 125 µm PEN [g] | infrared imaging | 3.58 | 4.3 | 1000 | 1.2 | [13] |

[a] $(Pb_{0.88}La_{0.08})(Zr_{0.65}Ti_{0.35})O_3$ [b] $Pb(Zr_{0.2}Ti_{0.8})O_3$ [c] $Ba_{0.67}Sr_{0.33}TiO_3$, nanowire array
[d] 68/32 mol% [e] 62.6/29.4/8 mol% [f] 62.6/29.4/8 mol%
[g] polyethylene naphthalate * not reported

## S2.) Structural and microstructural characterization

XRD patterns of the sample stack were recorded using a high-resolution diffractometer X'Pert Pro (PANalytical B.V., Netherland) with Cu-K$\alpha_1$ radiation (45 kV, 40 mA) in reflection mode. Diffraction patterns were recorded at room temperature (RT) in the range of 10–120° 2$\theta$ using a 1D detector (X'Celerator, PANalytical, Netherland) with a capture angle of 2.122°. The exposure time for each step was 100 s, and the interval between the obtained data points was 0.017°. The phase identification was performed using the PDF-4+ database. The microstructure and thickness of the prepared sample stack were examined using a field-emission scanning electron microscope (FE-SEM, JSM-7600F, Jeol Ltd., Japan). Prior to microstructural analysis, specimens were cut for cross-section examination, then ground, polished to 0.25 µm with diamond paste, and finely polished with a colloidal silica suspension (OP-S, Struers, Denmark) for polished-surface examination. SEM micrographs were taken with a backscattered-electron detector at an accelerating voltage of 15 kV and a working distance of 15 mm.

The XRD pattern of the PMN–10PT sample stack is shown in Fig. S1a. Only the peaks corresponding to the pseudocubic perovskite reflections and Au phase are present, and no additional secondary phase peaks, such as the commonly present $Pb_3Nb_4O_{13}$ pyrochlore phase, were observed. Fig. S1b shows SEM micrographs of polished cross-sectional areas of the sample stack. The micrographs show a polyimide substrate (black), a ~1.0 µm thick Au bottom electrode (light grey) and a ~3.0 µm thick PMN–10PT ceramic layer (dark grey) running from bottom to top in the top image in Fig. S1b. The thicknesses of the Au electrode and the ceramic layer were determined at different locations of the cross-section. The microstructure shows a uniform, homogeneous and dense ceramic film consisting of small grains and a small number of pores (further details can be found elsewhere).[14,15]

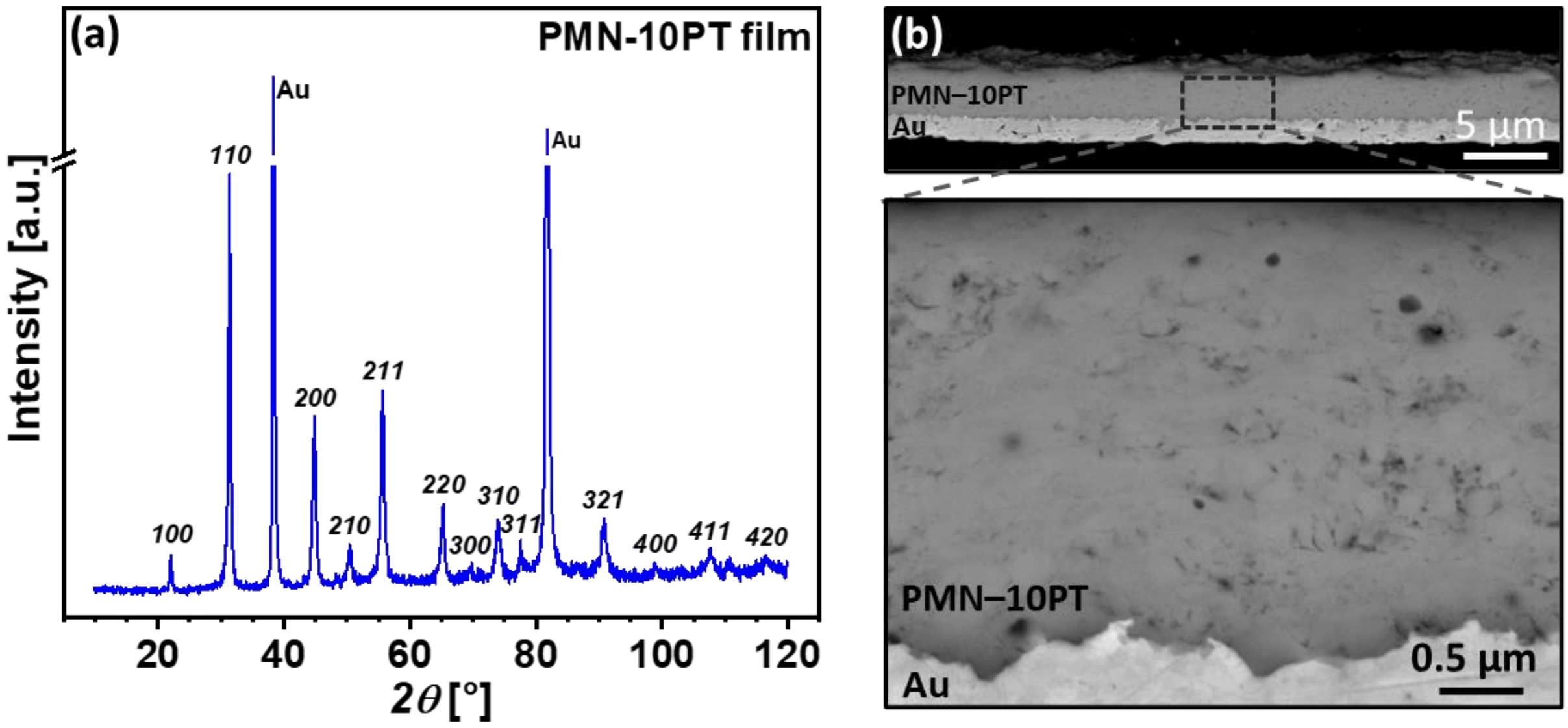

Figure S1. (a) Room-temperature XRD pattern (corresponding to Cu K$\alpha_1$ only) and (b) SEM micrographs of a polished cross-section of a PMN–10PT film on a polyimide substrate and a sputtered bottom Au electrode. The indexed peaks of the perovskite phase are labeled according to cubic notation.

## S3.) Dielectric and ferroelectric properties

Relative dielectric permittivity ($\varepsilon_r$) and dielectric loss (tan$\delta$) as a function of temperature were analyzed during cooling from 200 °C to 25 °C (1 K min$^{-1}$) using a modified tube furnace coupled with a HP 4284A (Hewlett-Packard, California, USA) Precision LCR impedance meter in the frequency range of 0.1–100 kHz. AC current density–electric field (*j*–*E*) loops were measured at RT using an Aixacct TF analyzer 2000E by applying a triangular waveform with a frequency of 1 kHz and electric fields up to 700 kV cm$^{-1}$. The polarization–electric field (*P*–*E*) hysteresis loops were determined simultaneously with the Aixacct software by numerical integration of the current signal.

The temperature and frequency ($\nu$) dependence of the real component of the relative permittivity ($\varepsilon_r'$) and tan$\delta$ of the PMN–10PT film are shown in Fig. S2a. The fabricated ceramic film exhibits a broad and diffuse peak-permittivity with a weak frequency dispersion reaching a maximum value of ~755 at 55 °C and 1 kHz, indicating a relaxor-like ferroelectric behavior. The sample exhibits relatively low dielectric loss over the entire measured temperature range, ranging from 0.7% at 25 °C to 6.2% at 200 °C (both at 1 kHz). The PMN–10PT film exhibits relatively slim and tilted relaxor-like *P*–*E* hysteresis loops (Fig. S2b) with a coercive electric field ($E_c$), remanent polarization ($P_r$), and maximum polarization ($P_{max}$) of ~37 kV cm$^{-1}$, ~5 μC cm$^{-2}$, and ~34 μC cm$^{-2}$, respectively (all values were determined at 25 °C, 700 kV cm$^{-1}$ and 1 kHz).

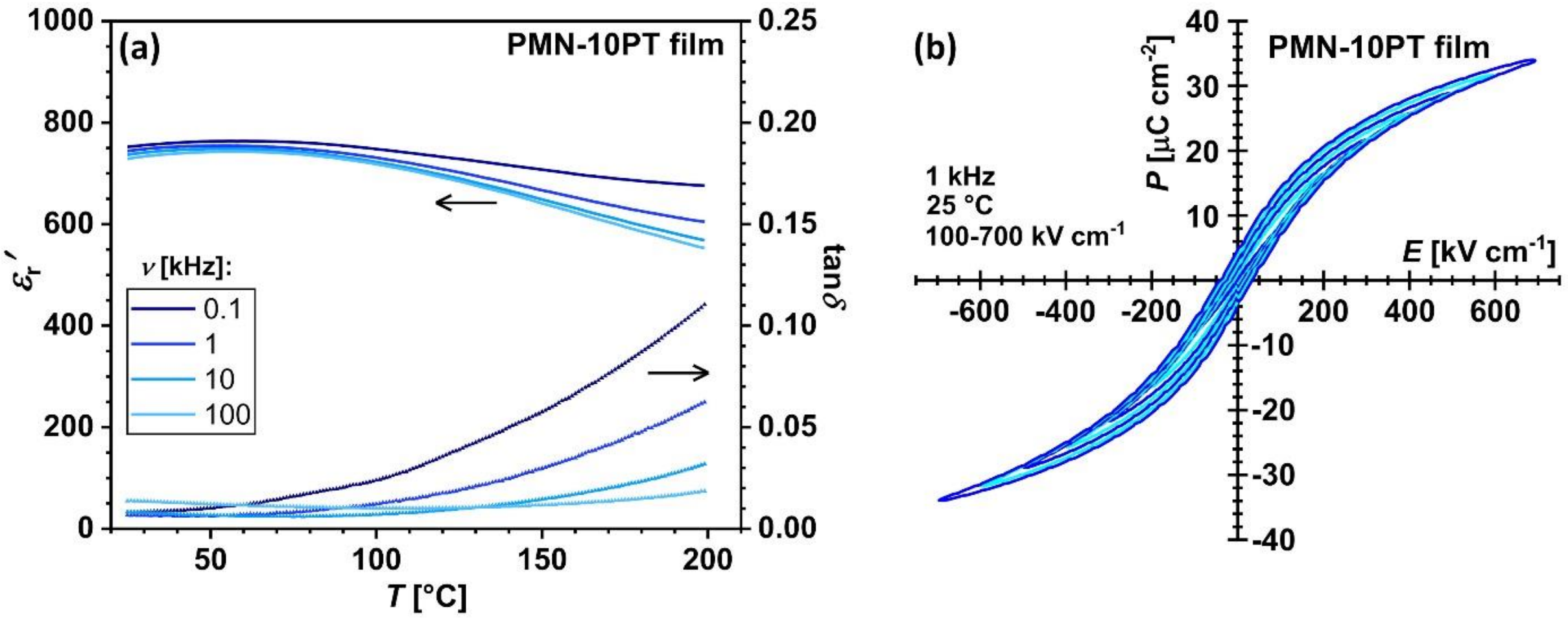

Figure S2. (a) Temperature and frequency dependence of $\varepsilon_r'$ and tan$\delta$ and (b) *P*–*E* hysteresis loops of a PMN–10PT film on a polyimide substrate measured at room temperature with increasing electric-field amplitude and a driving frequency of 1 kHz.

## S4.) Direct electrocaloric measurements with infrared camera

The measurement setup for the direct EC measurements with the IR camera is schematically shown in Fig. S3. The entire setup consists of a commercially available IR camera (FLIR X6580sc), an arbitrary waveform generator (Keithley 3390), a high-voltage amplifier (Trek 2220), and an oscilloscope (Agilent DSO5014A). Ambient temperature was controlled with a temperature stage (Linkam, THM S600). All instruments were controlled and monitored by a Python script, which also allowed fast data acquisition. For the EC measurements, the samples were placed directly on a temperature stage, while the electrical contacts were made with micromanipulators. The distance between the sample and the IR camera was set to 11 cm.

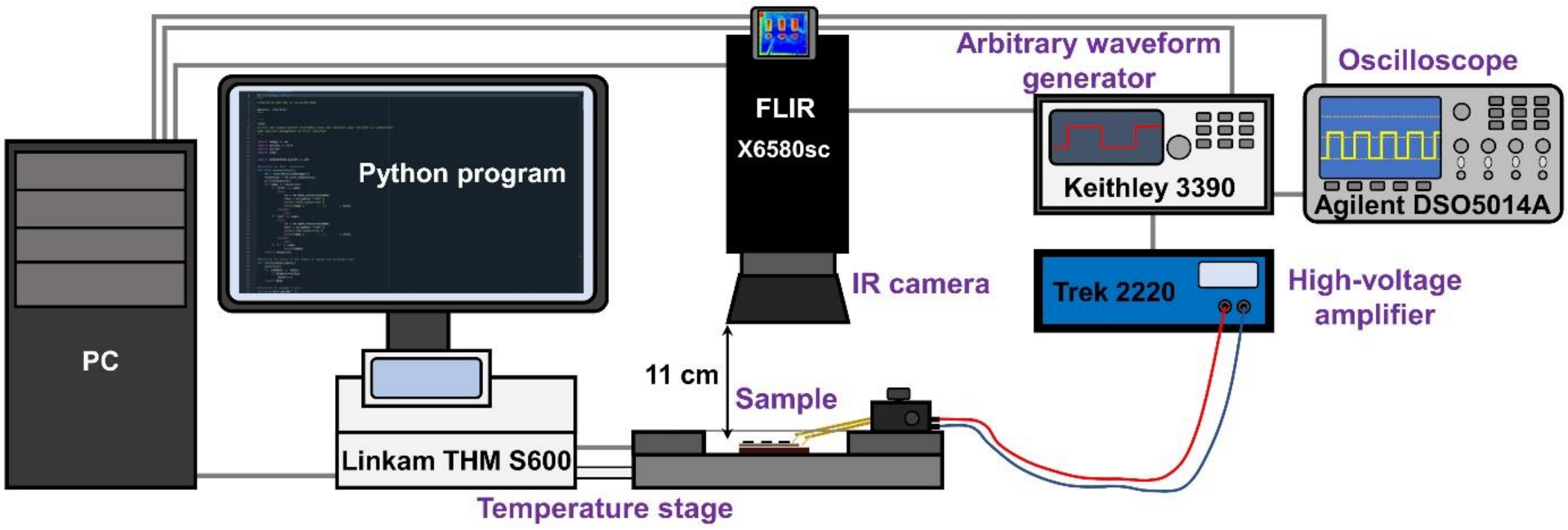

Figure S3. Schematic representation of the experimental setup for direct EC measurements with the infrared camera. Note that the elements are not shown at the same scale.

An example of the EC cycle typically recorded by the IR camera, consisting of adiabatic polarization (EC heating peak) and adiabatic depolarization (EC cooling peak), is shown in Fig. S4. The high frame rate of the IR camera (1.35 kHz) allows us to track the rapid heat transfer, also called internal thermal relaxation, between the EC active ceramic film and the substrate/bottom electrode stack. A high temporal resolution of ~0.74 ms allows us to capture a substantial part of the intrinsic EC effect of the ceramic film. The measured EC induced $\Delta T$ was obtained from the cooling peak that occurred ~1.5 ms after the electric field was turned off (inset in Fig. S4). The fast internal relaxation time allows us to reduce the on-field time to 0.5 s, minimizing the risk of dielectric breakdown. However, tracking the fast dynamics of a heat transfer resulted in a slightly larger measurement noise, which was smoothed by considering an average of several measured EC cycles. To further improve the noise of the measured temperature (visible as a wavy baseline in Fig. S4), an additional enclosure must be implemented to mimic a closed chamber furnace.

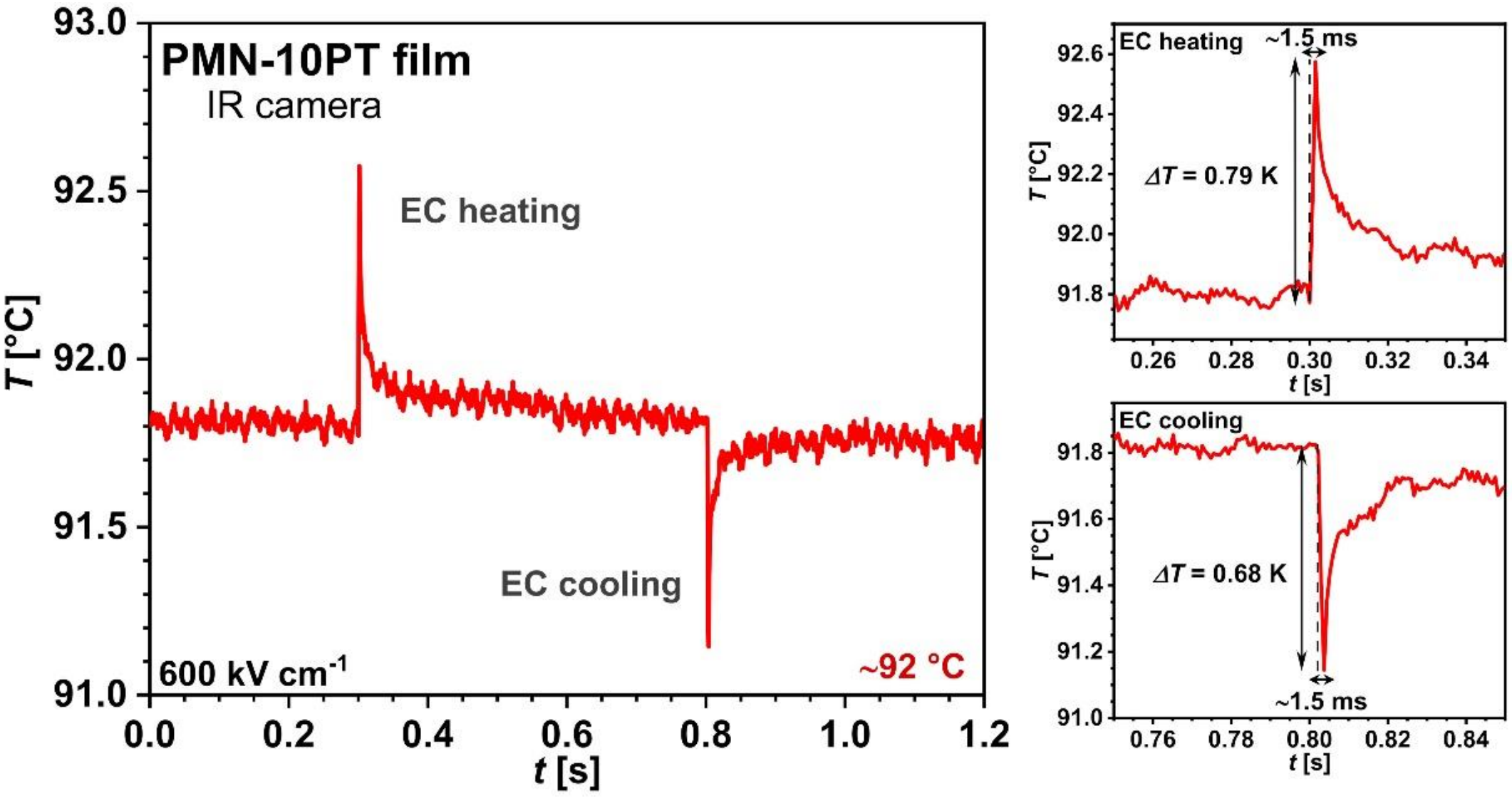

Figure S4. EC cycle measured with the infrared camera at ~92 °C and 600 kV cm$^{-1}$, with insets showing the magnified view of a selected heating and cooling peak.

IR thermogram of the top surface of the entire sample stack with the marked thick black ink areas (blue cross) and the printed black ink layers (red crosses) is shown in Fig. S5a. Assuming an emissivity of 1, the actual temperature of the sample was determined using the thick black ink regions, which was then used to calculate the emissivity of the thinner printed black ink layers. The importance of the additional black ink is clearly seen in the thermal image of the top Au electrode (Fig. S5a), where the top electrode always appears to be at RT (or sometimes even below), even though the set temperature of the heating stage was 50 °C. This is a consequence of the low emissivity (or, conversely, the high thermal reflectivity) of Au, which indicates the temperature of the laboratory environment (or sometimes the temperature of the cooled detector) rather than the temperature of the sample. One reason of the usage of a thinner black ink layer in addition to the thick black ink is the tradeoff between having a large enough emissivity to ensure a correct temperature measurement and a thin enough layer to minimize the additional EC inactive thermal mass on the sample surface. If the black ink layer is too thick, the measured $\Delta T$ will be reduced; if the layer is too thin, the surface emissivity will be too low. Note that for an accurate temperature measurement, the emissivity of the object must be greater than 0.5, and for an opaque object (i.e., with zero transmissivity), the sum of the emissivity and reflectivity is 1.

The temperature dependence of the emissivity of the printed black ink layers used for the EC measurements is shown in Fig. S5b. The emissivity reaches a value of ~0.92 at 30 °C and then drops rapidly to ~0.75 at 47 °C, while it stabilizes at ~0.70 at higher temperatures (up to 100 °C). The emissivity is reported as the mean and standard deviation of selected regions (marked with red crosses in Fig. S5a) and 10 different applied electric fields. The emissivity was calculated using the software provided by the manufacturer. It should be noted that the emissivity cannot be calculated correctly near RT because the contribution of the environment is too large and the determination error is therefore significant. Therefore, the emissivity could not be determined at temperatures below 30 °C and was therefore assumed to be 1.

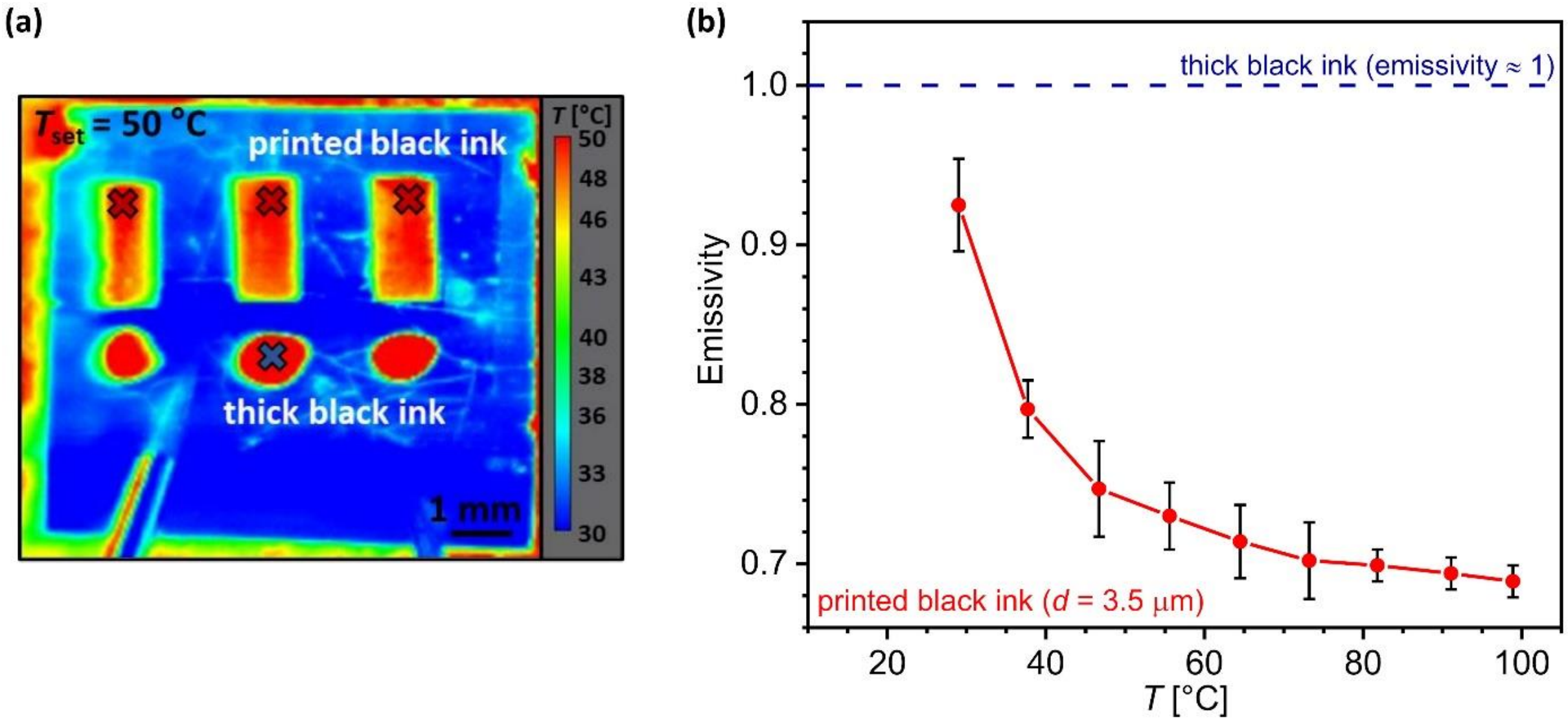

Figure S5. (a) Infrared thermogram of the sample used for the direct EC measurements. The area of thick black ink used to determine the actual temperature is marked with a blue cross, while the printed black ink layers used to determine the EC effect are marked with red crosses. (b) Temperature dependence of the emissivity of the printed black ink layers.

A commercial office printer was used to apply a thin layer of black ink, which proved to be a very convenient and inexpensive method of printing a relatively uniform layer of black ink with sufficient control over its thickness. The thickness of a printed layer used for the IR camera measurement was determined by profilometry (Fig. S6a). The profilometric measurement, performed on the exact same area used for the EC measurements, resulted in a fairly well-controlled thickness of ~(3.5 ± 0.5) μm, which was used for further finite element modeling (see SI, S5). Moreover, the thickness of the black ink layer was also confirmed by optical microscopy (Fig. S6b). In this case, the layer was printed on a polyethylene naphthalate (PEN) substrate ($d$ = 125 μm, Kaladex®, DuPont Teijin Films), which was precisely cut with a cryo-ultramicrotome (EM UC6, Leica Microsystems) to obtain the cross-section.

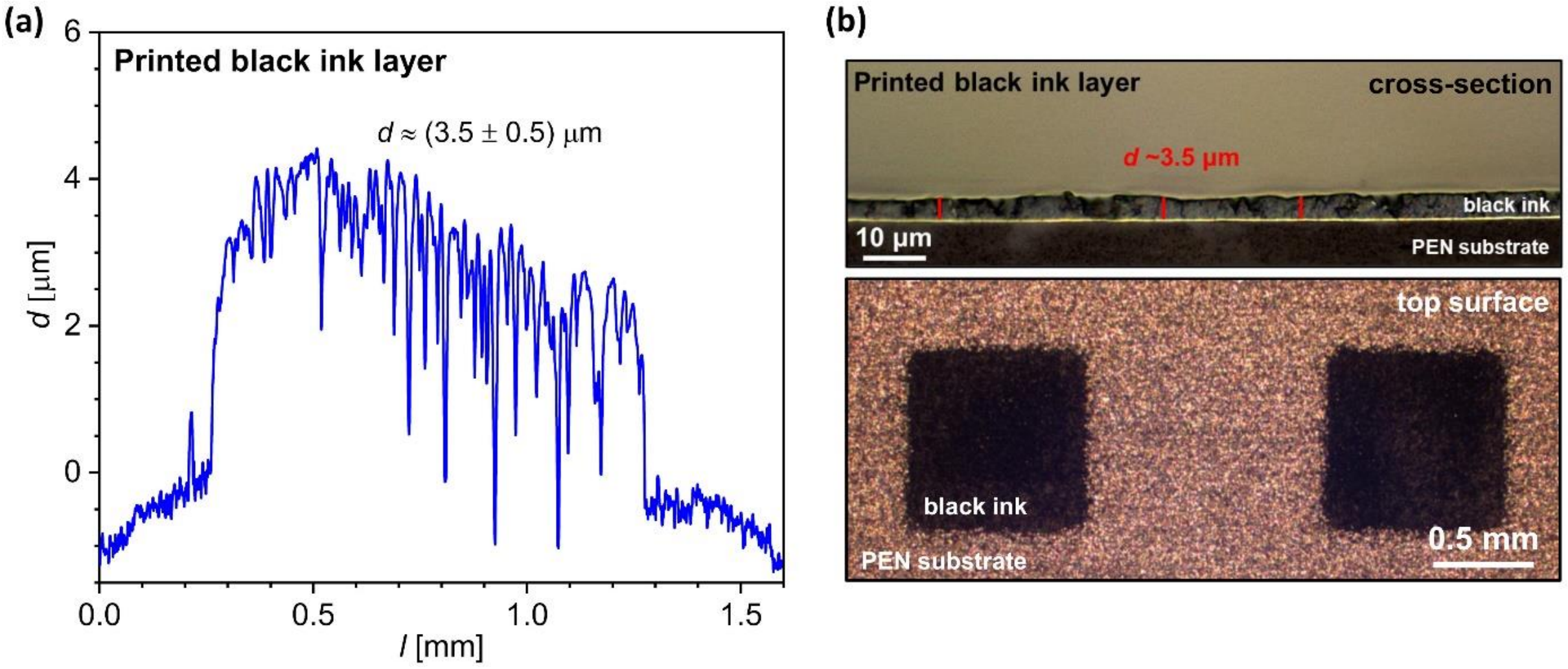

Figure S6. (a) Profilometric measurement and (b) optical microscope image of the cross-section and top view of the printed black ink layer.

## S5.) Finite element modeling

To determine the intrinsic $\Delta T_{EC}$ of the active ceramic film alone, a heat flow analysis of the entire sample stack was performed by the finite element method using COMSOL Multiphysics software. To construct a 2D model mimicking the sample stack used for the EC measurement, the real arrangement of the sample stack, the dimensions of its subcomponents, and their structural (density, $\rho$) and thermal parameters (specific heat capacity, $c_p$ and thermal conductivity, $\lambda$) were used. All details of the sample stack and the properties of its components are shown in Fig. S7. To simplify the model and reduce computational time, the top Au electrode was not included in the model because its thickness is an order of magnitude smaller compared to the other components. For a given sample stack (i.e., without the top electrode), a user-generated mesh is shown in Fig. S8.

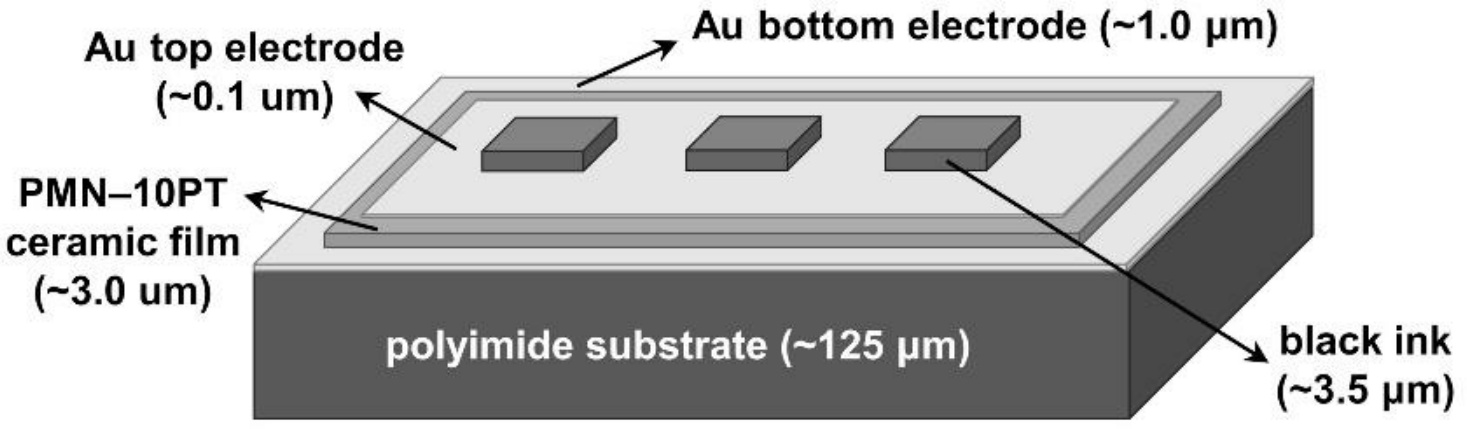

| Component | $d$ [µm] | $\rho$ [kg m$^{-3}$] | $c_p$ [J kg$^{-1}$K$^{-1}$] | $\lambda$ [W m$^{-1}$K$^{-1}$] |
|---|---|---|---|---|
| black ink | 3.5 | 1372 [a] | 710 [16] | 0.3 [16] |
| Au top electrode | 0.1 | 19300 [c] | 129 [17] | 318 [18] |
| PMN–10PT | 3.0 | 8130 [c] | 349 [a] | 1.27 [19] |
| Au bottom electrode | 1.0 | 19300 [c] | 129 [17] | 318 [18] |
| Polyimide (Kapton) * | 125 | 1454 [a] | 904 [a] | 0.12 [b] |
| $Al_2O_3$ | 125 | 3690 [d] | 880 [d] | 18 [d] |

* annealed at 400 °C [a] measured at 25 °C [b] Dupont™ Kapton® HN datasheet
[c] theoretical bulk density [d] COMSOL Multiphysics database

Figure S7. Schematic representation of the layout and dimensions of the entire sample stack used for the EC measurement, with a table summarizing the structural and thermal parameters of all components involved in the numerical model.

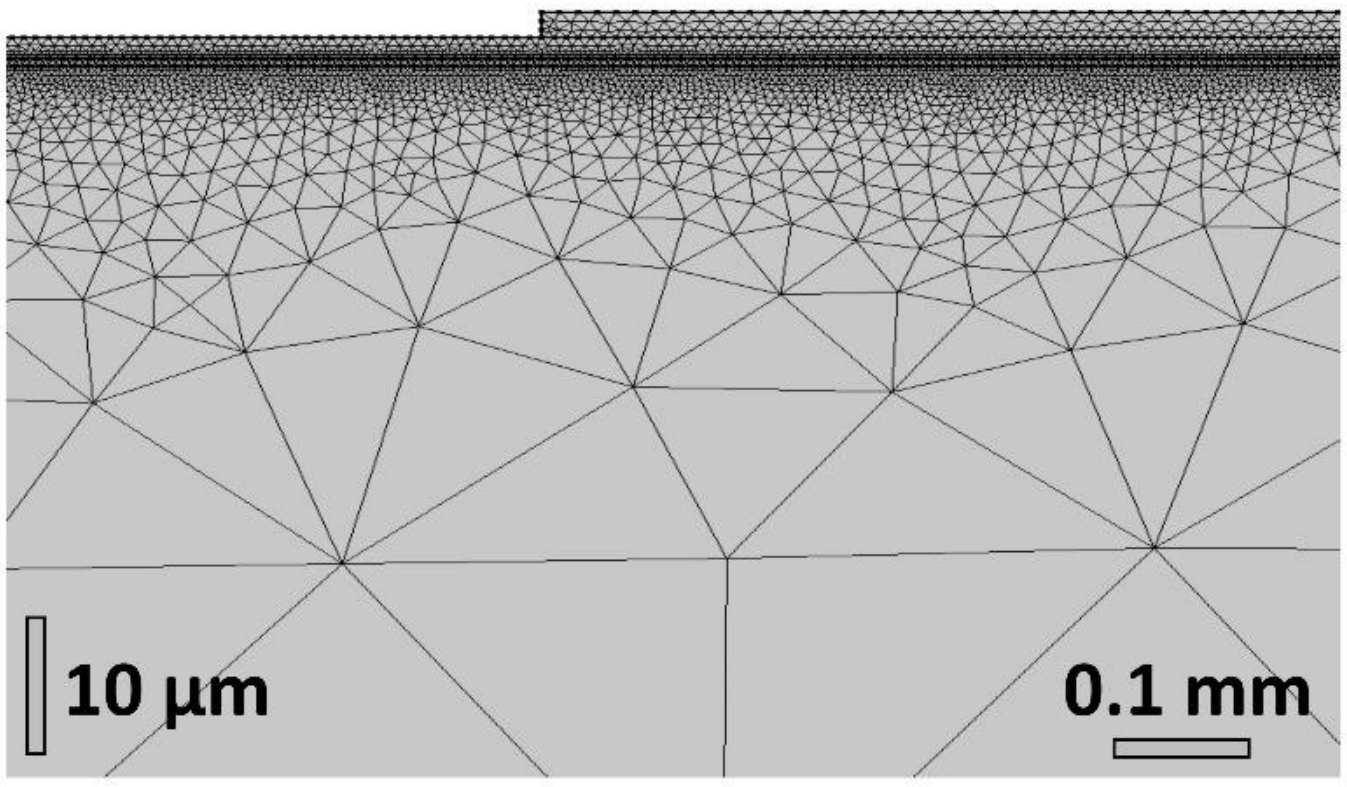

Figure S8. User-generated mesh used for modeling the intrinsic EC effect in the studied sample stack.

The correction factor $k$ was determined using a constructed finite element model that was refined and verified by fitting it to the experimental data. For simplicity, only positive heat pulses (EC heating peaks) were refined. The values of $k$ were determined by comparing the numerically determined $\Delta T_{ad.}$ inside the ceramic film and the measured $\Delta T_{meas.}$ on the top of the black ink layer for three different temperatures (i.e., ~22.5 °C, ~56 °C and ~92 °C) selected over the entire range of measured temperatures (i.e., 15–100 °C). The graphical representation of the constructed model fitted to the experimental data is shown in Fig. 3a, while the numerical values are listed in the table in Fig. S9. The correction factor obtained by comparing the ratio of modelled and measured $\Delta T$ ranges from 3.1 at ~22.5 °C to 3.5 at ~92 °C. The temperature dependence of $k$ was determined by combining the above results for three selected temperatures and interpolated values for intermediate temperatures (Fig. S9). In addition, $k$ for the two final temperatures (outside the range of the three selected temperatures) was assumed to be the same value as for the closest temperatures, namely 3.1 at ~15 °C and 3.5 at ~100 °C.

| $T$ [°C] | $\Delta T_{meas.}$ [K] | $\Delta T_{ad.}$ [K] | $k$ |
|---|---|---|---|
| 22.4 | 0.45 | 1.40 | 3.1 |
| 55.8 | 0.61 | 2.15 | 3.5 |
| 91.8 | 0.77 | 2.70 | 3.5 |

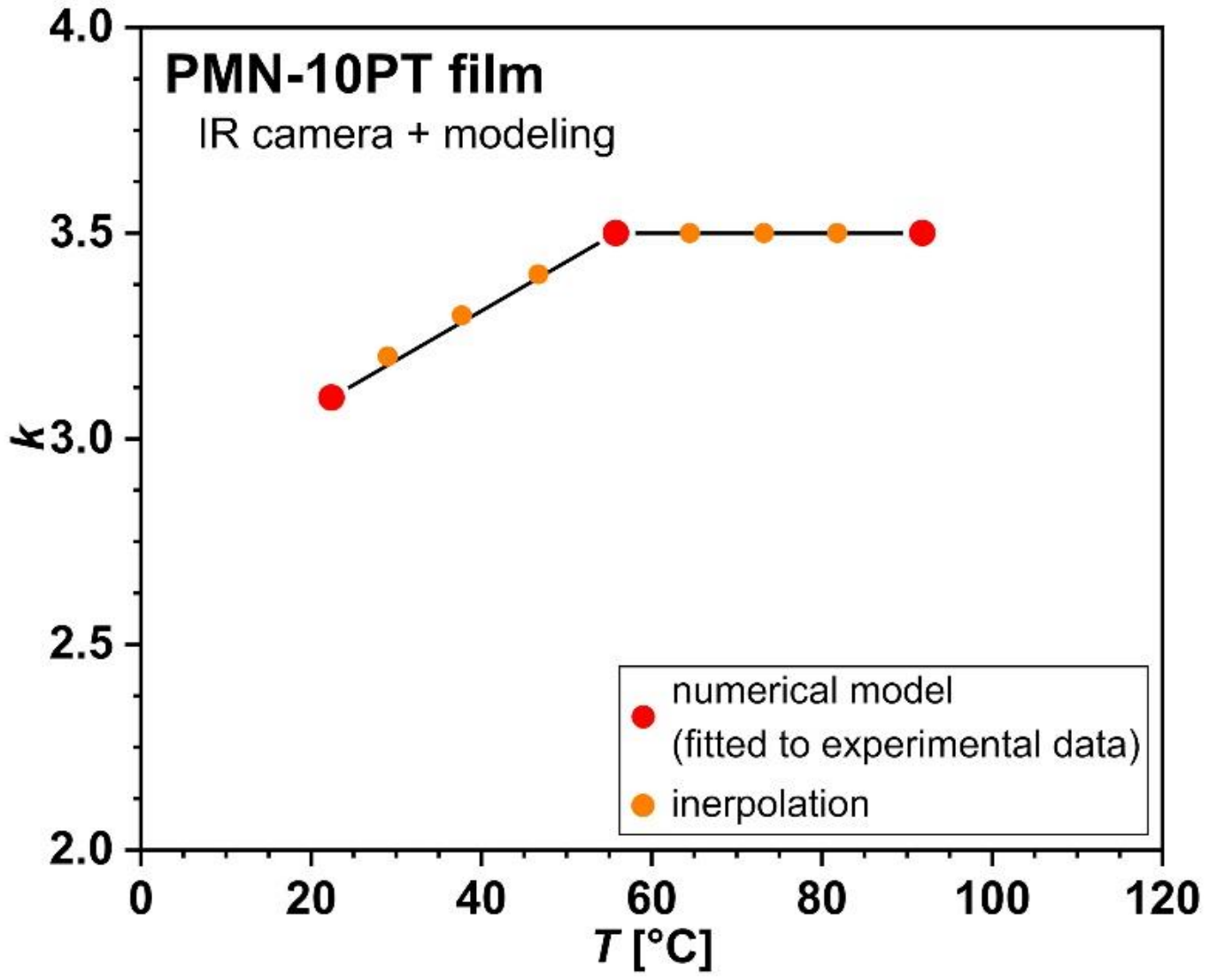

Figure S9. Temperature dependence of correction factor ($k$) determined by a numerical model fitted to the experimental data (table) and the interpolated values.

## S6.) Direct electrocaloric measurements with thermistor-in-calorimeter method

The EC effect of the studied films was additionally confirmed by another direct thermometric measurement method originally developed by Z. Kutnjak *et al*. and referred to in the literature as a modified high-resolution calorimeter with a small bead thermistor.[2,10,20] Some details of our in comparison with that of Z. Kutnjak *et al*.[2,10,20] slightly modified measuring method and setup, here called thermistor-in-calorimeter method, can also be found in Ref. [21]. In this study, the measurements were performed on the same samples used for the IR imaging (Fig. 2a). Note that no additional black ink layer is required for this measurement. The entire measurement setup is shown in Fig. S10a. The measurements were performed in a modified differential scanning calorimeter (DSC, Netzsch DSC 204 F1, Germany), which was used as a temperature-controlled unit and provided a precise temperature stabilization of ±1 mK. The EC induced $\Delta T$ was measured with a miniature radial glass thermistor (GR500KM4261J15, Measurement Specialties) glued to the top electrode of the sample (Fig. S10b). The resistance and thus the temperature of the thermistor were monitored using a Keithley 2100 digital multimeter (Keithley Instruments, Ohio, USA). The electric field to trigger the EC cycle was applied using a Keithley High Voltage Source-Measure Unit 237. In addition, a Pt1000 temperature sensor was used to control the ambient temperature in the DSC measurement cell. A LabVIEW program was developed for fast, automatic operation and data acquisition. To minimize the thermal connection of a sample stack to the holder and the measuring cell, and thus keep the EC induced heat in the sample longer, the entire sample stack was kept in the air by suspending it from the holder via electrical contacts and thermistor wires (Fig. S10).

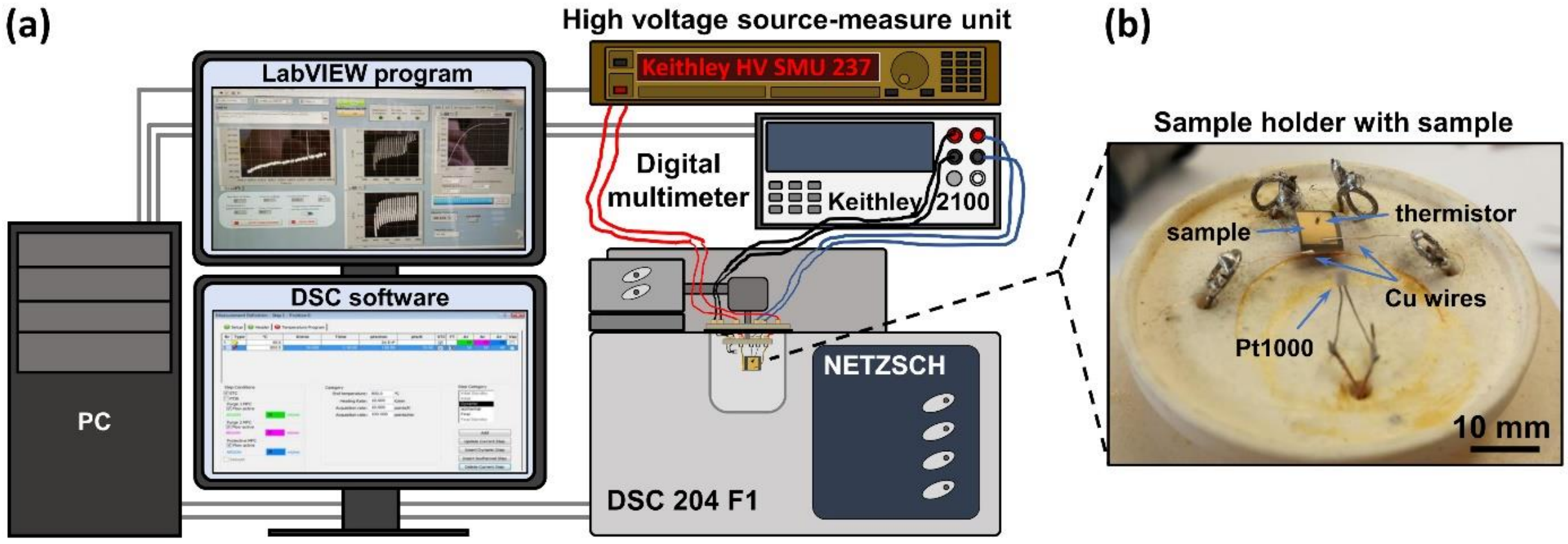

Figure S10. (a) Schematic representation of experimental setup for direct EC measurements using a thermistor-in-calorimeter method (note that elements are not shown to the same scale) and (b) photograph of a sample holder with attached sample prepared for EC measurements.

To allow relaxation of the EC induced $\Delta T$ to ambient temperature and thus facilitate determination of the $\Delta T_{EC}$, an electrical step signal with a period of 200 s (i.e., 100 s on-field and 100 s off-field) was used. The typically obtained EC cycles, consisting of adiabatic polarization (i.e., heating peak) and adiabatic depolarization (i.e., cooling peak), are shown in Fig. S11. EC measurements were performed in the temperature range of 27–105 °C and at electric field amplitudes of 100–400 kV cm$^{-1}$. For each selected ambient temperature, the temperature was first stabilized, followed by measurements at progressively applied electric fields before moving to the next temperature. Examples of EC cycles at various selected electric fields and temperatures are shown in Fig. S11a.

Because of the relatively long thermalization time of the thermistor (>0.7 s, as shown in Fig. S11a), the $\Delta T$ measured by the thermistor is much lower than the intrinsic EC effect of the ceramic film. This prevents us from tracking the internal thermal relaxation of the EC induced heat in the ceramic film to the surrounding EC non-active parts of the sample stack (mainly the substrate in this case) and thus capturing the essential part of the intrinsic EC effect of the film. However, the intrinsic EC effect can be determined from the so-called external thermal relaxation of the entire thermalized sample assembly (i.e., the entire sample stack with mounted wires and thermistor), which shows up as an exponential decrease/increase in the temperature peak back to the initial ambient temperature (visible as the baseline in the inset of Fig. S11a).

Assuming rapid internal thermal relaxation (i.e., there is no temperature gradient in the sample stack), the entire thermally balanced system slowly releases heat to the environment on a much longer time scale. This allows us to perform a simplified data analysis to fit the external thermal relaxation and obtain the EC induced $\Delta T$ of the entire sample assembly. A simple zero-dimensional model of the thermal system used to fit the experimental data is defined as follows:

$$T(t) = T_{amb} + \Delta T e^{-\frac{t}{\tau_{ext}}}, \qquad (1)$$

where $T_{\text{amb}}$ is the ambient temperature (baseline), $\Delta T$ is the EC temperature change of the entire internally equilibrated sample assembly (not $\Delta T$ of the ceramic film alone!), $t$ is the time after application/removal of the electric field, and $\tau_{\text{ext}}$ is the time constant of the external thermal relaxation process. More details on data analysis can be found in Refs. [2, 10]. An example of a curve fitted to the experimental data is shown in the inset of Fig. S11a. In this work, the $\Delta T$ obtained from fit is ~10% larger than the $\Delta T$ acquired directly from the thermistor, which corresponds to the heat losses to the environment within the thermalization time of the thermistor. The temperature- and field-dependent $\Delta T$ cooling peak values of the sample assembly after the fit are shown in Fig. S11b, where $\Delta T$ of ~24 mK was obtained at 27 °C and 400 kV $cm^{-1}$, while the maximum of ~37 mK was obtained at 105 °C and 380 kV $cm^{-1}$.

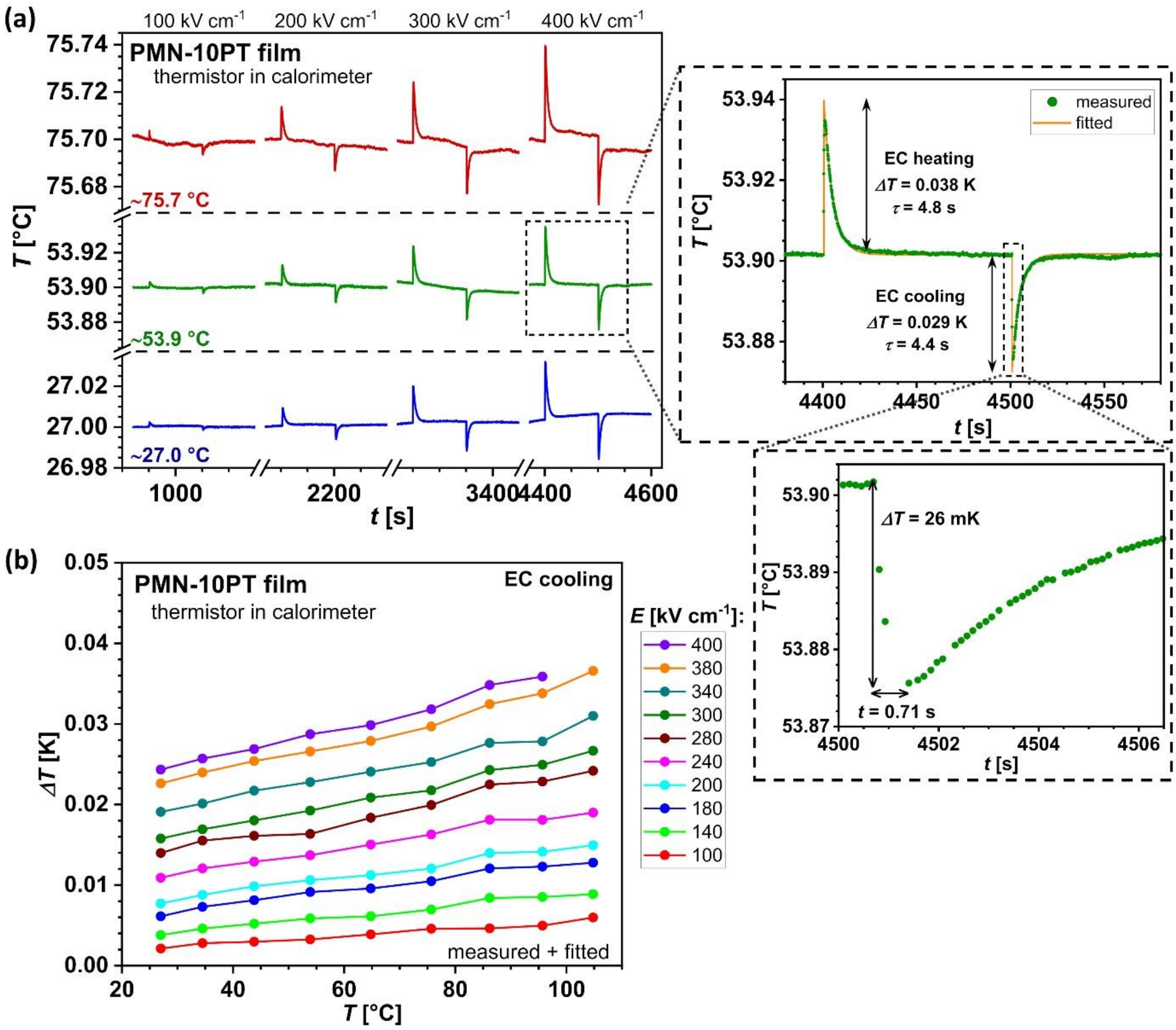

Figure S11. (a) Sequence of four EC cycles measured with the thermistor at different ambient temperatures and electric fields, with insets showing the magnified view of a selected heating and cooling peak with fitted zero-dimensional model function. (b) Temperature- and field-dependent $\Delta T$ corresponding to the measured and fitted cooling peaks of the entire sample assembly.

Once we obtain the $\Delta T$ of the entire thermalized sample assembly, we can derive the intrinsic $\Delta T_{EC}$ of the active film by considering the exact geometry of the sample stack and the specific heat capacities of all components involved in the measurement system. The overall sample assembly considered in the measurement of $\Delta T$ includes the entire sample stack (i.e., substrate, ceramic film, bottom and top electrodes), the thermistor, the wires for the electrical contacts, and the adhesive used to attach them to the sample surface. The structure of the sample with dimensions is shown schematically in Fig. S12a. If we know the heat capacities of all involved components (see table in Fig. S12b), we can determine $\Delta T_{EC}$ of the active film using the following equation:

$$\Delta T_{EC} = \frac{\sum_i C_p^i}{C_p^{EC}} \cdot \Delta T = k \cdot \Delta T, \tag{2}$$

where the sum of $C_p^i$ is the total heat capacity of the system (i.e., EC active and non-active components) and $C_p^{EC}$ is the heat capacity of only the EC active (i.e., electroded) part of the sample stack. The ratio between the total heat capacity of the system and the EC active part of the sample gives a correction factor ($k$) needed to obtain the intrinsic $\Delta T_{EC}$ of the active material. As expected, comparison of all contributions to the total heat capacity of the system shows that the substrate makes the largest contribution (i.e., ~84% at 25 °C) to the magnitude of the correction factor (Fig. S12b). To obtain more accurate values for correction factor over the entire range of measured temperatures, the temperature dependence of $k$ was determined by considering the change in specific heat capacity ($c_p$) of the polyimide substrate alone, since it makes the largest contribution to the total $k$ (Fig. S13b). This results in an increasing trend of $k$ with increasing temperature, reaching a value of ~40 at RT and ~50 at 100 °C (Figs. S13a and S13b).

The temperature dependence of the $c_p$ of the polyimide substrate after annealing at 400 °C (which corresponds exactly to the conditions used for the measured sample) was determined from differential scanning calorimetry curves measured with the same DSC (in this case in its original differential scanning mode). Measurements were performed on disc-shaped plates ($2r$ ~6 mm, $d$ = 0.125 mm) placed in a Pt crucible with a lid. To exclude possible thermal effects related to evaporation of water and desorption of gasses from the surface of the sample, two heating and cooling cycles were performed. In the first cycle, the samples were heated from 25 °C to 120 °C and then cooled to 0 °C with liquid nitrogen. Subsequently, the samples were heated again to 120 °C and cooled again to 25 °C. Heating and cooling rates of 5 K min$^{-1}$ were used for all measurements. Sapphire (Netzsch, $2r$ = 5.2 mm, $d$ = 0.25 mm) was used as the standard material to determine the $c_p$ of the sample. The $c_p(T)$ values shown in Fig. S13b were determined from the second heating DSC run considering the known $c_p$ values of the standard material.[22]

(a)

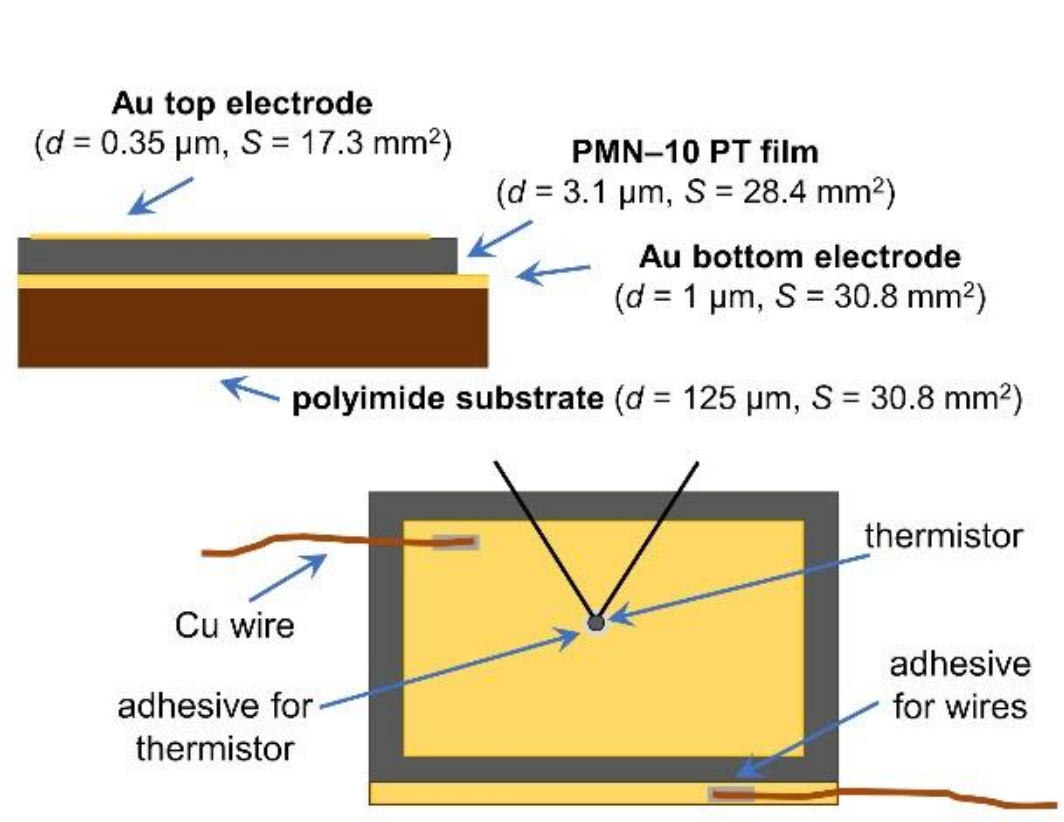

(b)

| No. | Component | Heat capacity [$10^{-3}$ J K$^{-1}$] | Percentage [%] |
|---|---|---|---|
| 1 | Cu wires | 0.283 | 4.7 |
| 2 | Thermistor | 0.079 | 1.3 |
| 3 | Adhesive for Cu wires | 0.074 | 1.2 |
| 4 | Adhesive for thermistor | 0.209 | 3.5 |
| 5 | Top Au electrode | 0.015 | 0.2 |
| 6 | Bottom Au electrode | 0.077 | 1.3 |
| 7 | PMN−10PT (electroded part) | 0.152 | 2.5 |
| 8 | PMN−10PT (non-electroded part) | 0.098 | 1.6 |
| 9 | Substrate | 5.057 | 83.7 |
| | **total** | **6.04** | **100.0** |

**Contributions to the total heat capacity:**

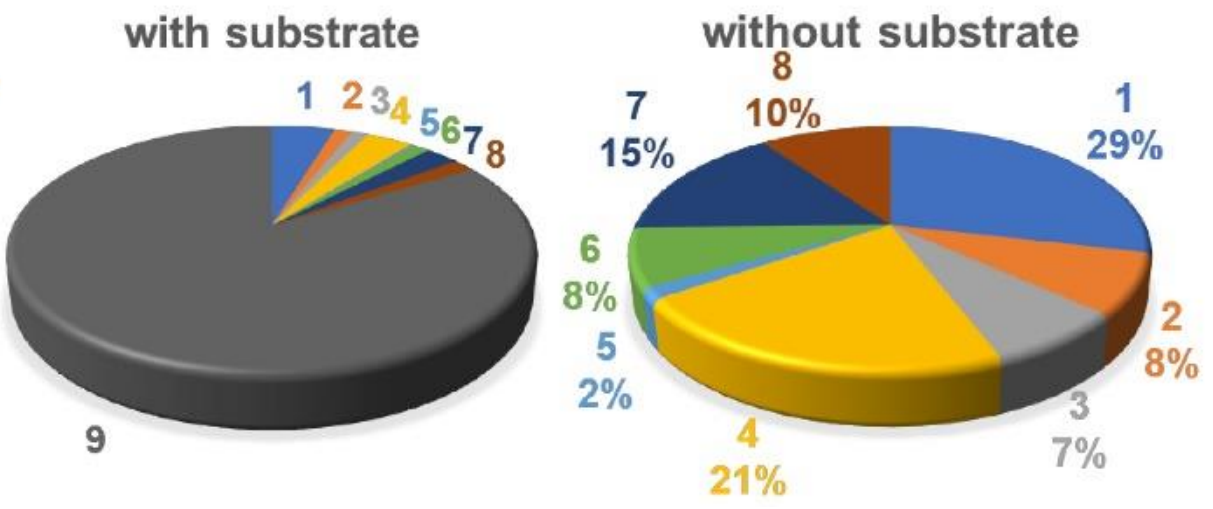

Figure S12. (a) Schematic representation of the entire sample assembly and its geometry for direct EC measurement by thermistor-in-calorimeter method. Note that the elements are not shown at the same scale. (b) Contributions of the individual components to the total heat capacity of the entire sample assembly required to obtain the correction factor at 25 °C.

(a)

| No. | Component | Heat capacity [$10^{-3}$ J $K^{-1}$] | | | |
|---|---|---|---|---|---|
| | | 25 °C | 50 °C | 75 °C | 100 °C |
| 1 | Cu wires | 0.283 | | | |
| 2 | thermistor | 0.079 | | | |
| 3 | adhesive for Cu wires | 0.074 | | | |
| 4 | adhesive for thermistor | 0.209 | | | |
| 5 | top Au electrode | 0.015 | | | |
| 6 | bottom Au electrode | 0.077 | | | |
| 7 | PMN−10PT (electoded part) | 0.152 | | | |
| 8 | PMN−10PT (non-electoded part) | 0.098 | | | |
| 9 | substrate | 5.057 | 5.480 | 5.984 | 6.588 |
| | **sum** | **6.04** | **6.47** | **6.97** | **7.58** |
| | **correction factor** | **39.7** | **42.4** | **45.7** | **49.7** |

(b)

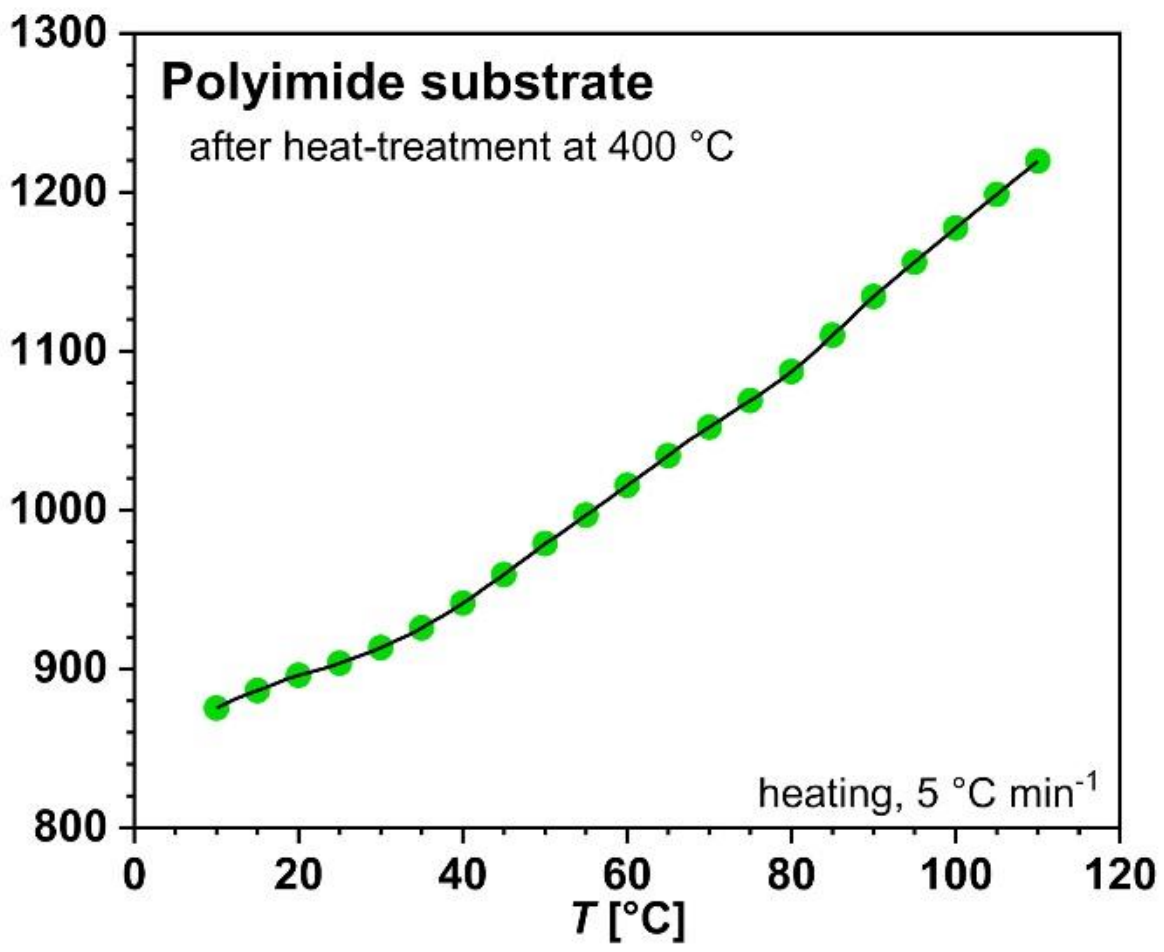

(c)

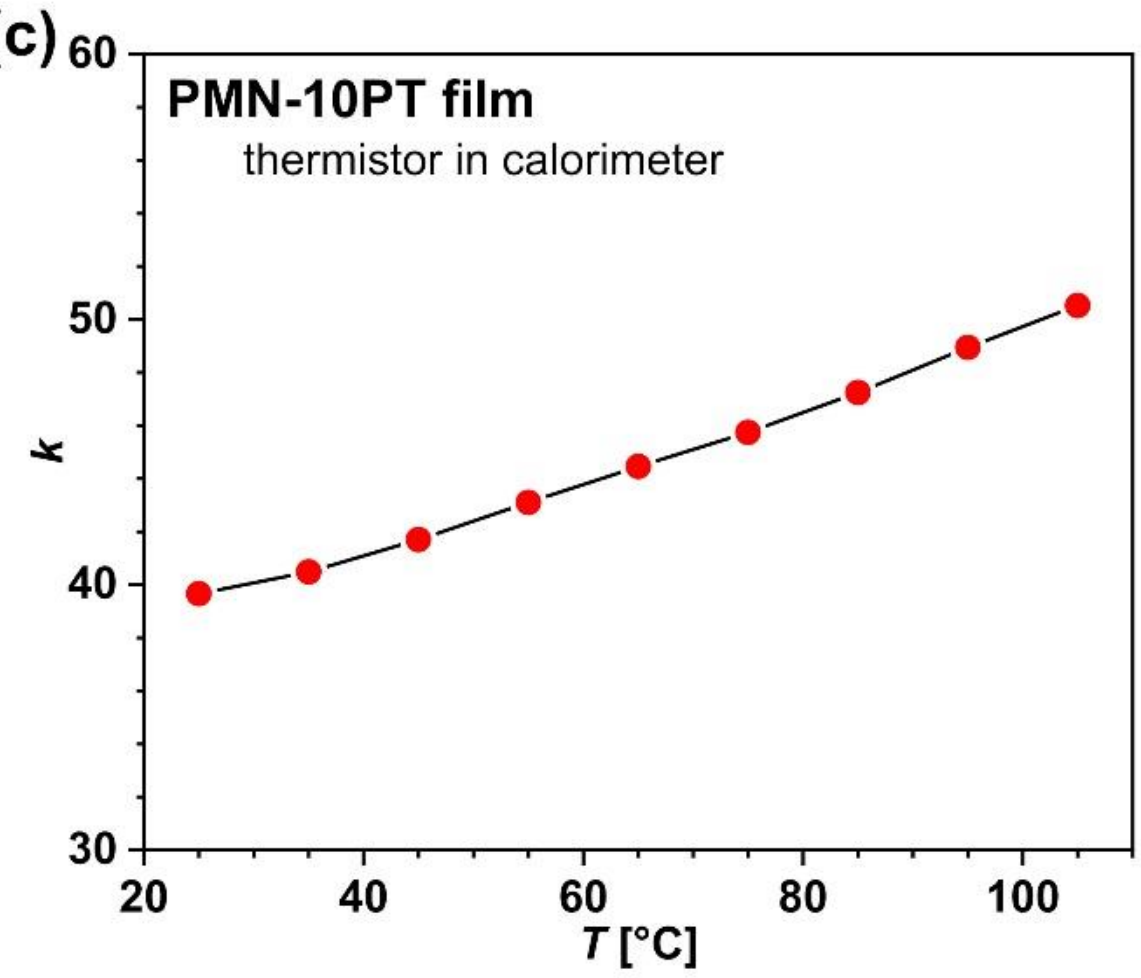

Figure S13. (a) Total heat capacities and corresponding correction factors of the entire sample assembly at different temperatures when (b) temperature dependence of $c_p$ for polyimide substrate was considered. (c) Temperature dependence of correction factor used to determine the $\Delta T_{EC}$.

Finally, combining the $\Delta T$ of the entire thermalized sample assembly (Fig. S11b) with a correction factor derived from the ratio of the heat capacities of all involved components (Fig. S13a), we can derive the intrinsic $\Delta T_{EC}$ of the PMN–10PT ceramic film (Fig. S14), resulting in a $\Delta T_{EC}$ of ~0.97 K at 27 °C and 400 kV $cm^{-1}$, while the maximum of ~1.85 K is reached at 105 °C and 380 kV $cm^{-1}$.

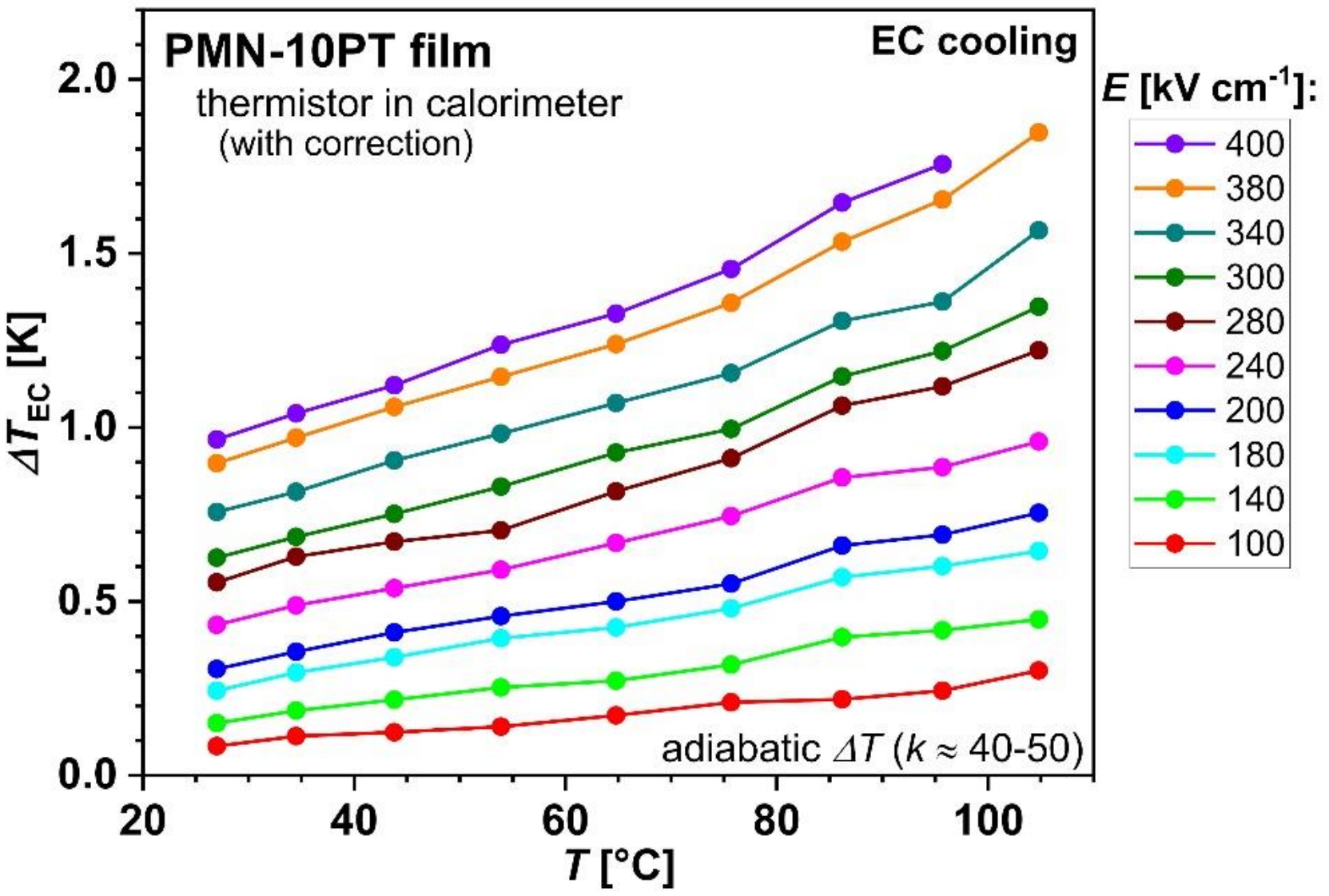

Figure S14. Temperature- and field-dependent $\Delta T_{EC}$ of PMN–10PT film determined by thermistor-in-calorimeter method and temperature-dependent correction factor.

In summary, the presented direct EC method accounts for temperature variations throughout the entire sample stack (including the substrate) and therefore requires a large correction factor to account for intrinsic $\Delta T_{EC}$ in the active film alone. Despite the somewhat larger correction, the method, if properly applied, can successfully account for the EC effect even in ceramic films on the substrate. However, one should be careful with the correction factor. Although it plays a crucial role in estimating the intrinsic EC effect of materials, it is rarely reported in the literature. Therefore, it is important to describe its determination in detail to avoid overestimation of the final $\Delta T_{EC}$ and thus maintain confidence in the method. Since the value of $k$ in ceramic films can easily exceed 100, it is even more important to specify it correctly to avoid large errors in determining the final EC effect.

The main requirements of the method, i.e., (i) a fast internal thermal relaxation time to ensure thermal equilibrium, and (ii) a sufficiently large ratio between EC active and non-active part of the sample assembly that prevents all EC induced heat from being lost before thermistor thermalization occurs, could be the main limitation in further reducing the thickness of the studied films. If the thermal mass of the active part of the sample is too small compared to the surrounding components (e.g., in the case of sub-micron thick ceramic films), the EC induced $\Delta T$ of the whole sample stack may either be negligible (i.e., it cannot be detected by the thermistor after its thermalization) or it may be very small and require a correction factor of several 1000, which may lead to a large error in the determination of the final intrinsic EC effect.

## S7.) Indirect electrocaloric measurements - Maxwell method

The electrocaloric effect is often determined by indirect EC measurement methods that provide a quick estimate of the magnitude of the EC effect by simply considering the ferroelectric and/or dielectric properties of the material. The EC effect is a thermodynamic phenomenon that can be described by thermodynamic laws.[2,20,23,24] One of the most commonly used approaches derives the EC effect from the Gibbs free energy density, into which one of Maxwell's relations describing the relationship between $\left(\frac{\partial P}{\partial T}\right)_E$ and $\left(\frac{\partial S}{\partial E}\right)_T$ can be implemented, leading to the well-known equation for isothermal EC entropy change ($\Delta S_{\mathrm{EC}}$):

$$\Delta S_{\mathrm{EC}} = \int_{E_i}^{E_f} \left(\frac{\partial P}{\partial T}\right)_E dE, \tag{3}$$

where $P$ is the polarization, $T$ is the temperature, $E_{\mathrm{i}}$ and $E_{\mathrm{f}}$ are the initial and final electric fields, respectively. On the other hand, the adiabatic EC temperature change ($\Delta T_{\mathrm{EC}}$) can be described as:

$$\Delta T_{\mathrm{EC}} = -\int_{E_i}^{E_f} \frac{T}{\rho\, c_p} \left(\frac{\partial P}{\partial T}\right)_E dE, \tag{4}$$

where $\rho$ is the density of the material and $c_{\mathrm{p}}$ is the specific heat capacity at a given constant electric field. Since both equations are based on the Maxwell relation between the ($S$, $T$) and ($P$, $E$) pairs of parameters, the method is commonly referred to as the Maxwell indirect method. Based on equations (3) and (4), the EC effect can be predicted by simply measuring the temperature- and field-dependent polarization data, $P(T, E)$ and $c_{\mathrm{p}}(T, E)$ of the material, assuming that there are no singularities in both data, such as the presence of a first-order phase transition. In addition, equation (4) is often simplified by considering electric-filed independent $c_{\mathrm{p}}(T)$ data, resulting in the $c_{\mathrm{p}}(T)$ term being outside the integral, which results in simplified integration of the equation (4).[2,20,23,24]

Although the method has proven to work well for the ergodic systems (i.e., ferroelectrics), the validity of the method used for non-ergodic systems, such as relaxor ferroelectrics (relaxors for short), is still unclear. Relaxors exhibit chemical and compositional perturbations that disrupt the long-range ferroelectric order. Consequently, relaxors are not in thermal or mechanical equilibrium, so the assumption of ergodicity in the Maxwell relation no longer holds. Therefore, this approach may not provide reliable results for relaxors and should be used with caution.[2,20,23,24]

The $P(T, E)$ data used to calculate the EC effect using equations (3) and (4) was obtained from measurements of $P$–$E$ hysteresis loops at various temperatures. $P$–$E$ hysteresis loops were determined by simultaneous numerical integration of the current signal loops measured with the Aixacct TF analyzer 2000E. The $P$–$E$ loops were determined by applying a unipolar and bipolar triangular waveform with a frequency of 1 kHz and electric fields from 100 to 400 kV cm$^{-1}$ with a step of 50 kV cm$^{-1}$. The measurements were performed in a temperature range of 10–110 °C, with a step size of 10 °C.

To calculate the derivative in equations (3) and (4), the results of both unipolar and bipolar $P$–$E$ loops were combined. The temperature and electric field dependence of $\Delta P$ was obtained from the upper branch of the unipolar loops (Fig. S15a), whereas the temperature dependence of the remanent polarization ($P_R$) was obtained from the bipolar loops at maximum applied electric field of 400 kV cm$^{-1}$ (Fig. S15b). To obtain the $P(T, E)$ values used for the calculations, the $\Delta P$ and $P_R$ data (Fig. S15c) were summed and fitted by quadratic polynomial regression (Fig. S15d) to obtain a smooth function, which was then used for integration. It should be noted that the values of $P_R$ in relaxors are usually much smaller compared to the $\Delta P$ part and therefore have little effect on the total $P$ value and its temperature-dependent trend. In ferroelectrics, however, the $P_R$ part plays an important role, as $P_R$ can significantly affect the magnitude and slope of the $P(T)$ data and thus the final $\Delta T_{EC}$(T) values.

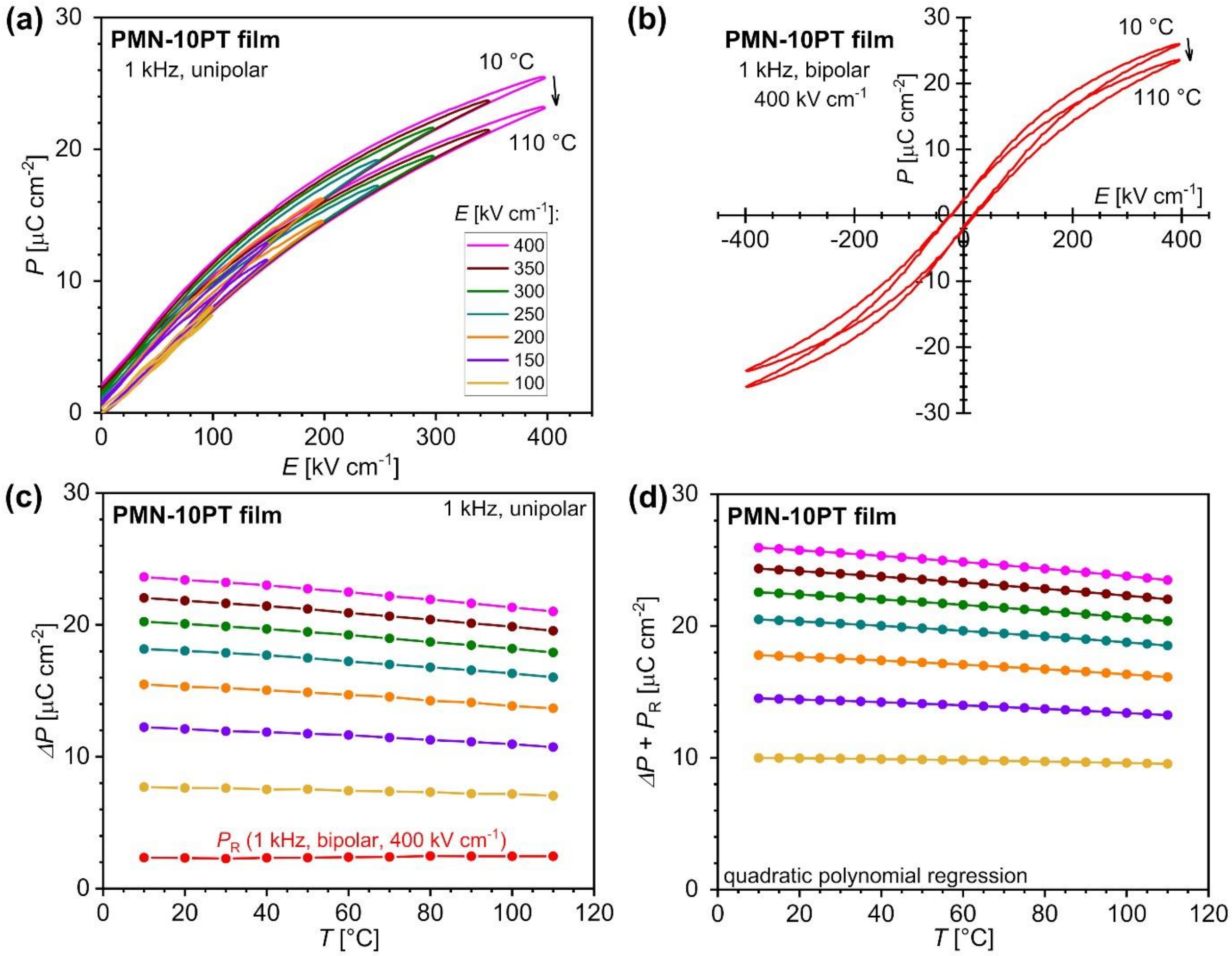

Figure S15. (a) Unipolar and (b) bipolar $P$–$E$ loops measured at different applied electric fields and temperatures at 1 kHz. (c) Temperature and field dependence of $\Delta P$ obtained from unipolar $P$–$E$ loops (upper branch) and $P_R$ obtained from bipolar $P$–$E$ loops at 400 kV cm$^{-1}$. (d) Temperature and field dependence of $\Delta P + P_R$ (obtained from unipolar and bipolar loops, respectively) after quadratic polynomial regression.

When deriving the EC effect from the $P$–$E$ loops, the shape and magnitude of the polarization data have a large effect on the final calculated EC effect. Therefore, a correctly measured polarization, i.e., the one representing the actual ferroelectric/dielectric properties of the material, is a key element for a reliable estimate of the EC effect. It should be noted that the EC effect can in principle be calculated from any $P$–$E$ hysteresis loop, whether it represents the actual polarization values or those with a large contribution from a leakage current. Consequently, in many studies the indirectly determined EC effect is calculated from *P*–*E* loops whose shape and size do not correspond to the intrinsic properties of the material, resulting in incorrectly estimated EC effects. The influence of frequency on the shape of the *P*–*E* loop is also clearly visible in the case of the prepared PMN–10PT films, where rounder and leakier loops were obtained at a lower frequency of 0.1 kHz than at 1 kHz and higher temperatures (Fig. S16). Consequently, the $\Delta T_{EC}$ values obtained from the *P*–*E* loops measured at 0.1 kHz gave different values and a temperature-dependent trend than the values obtained from the *P*–*E* loops measured at 1 kHz (not shown here). This once again underlines the importance of correctly measured polarization data for the derived EC effect values.

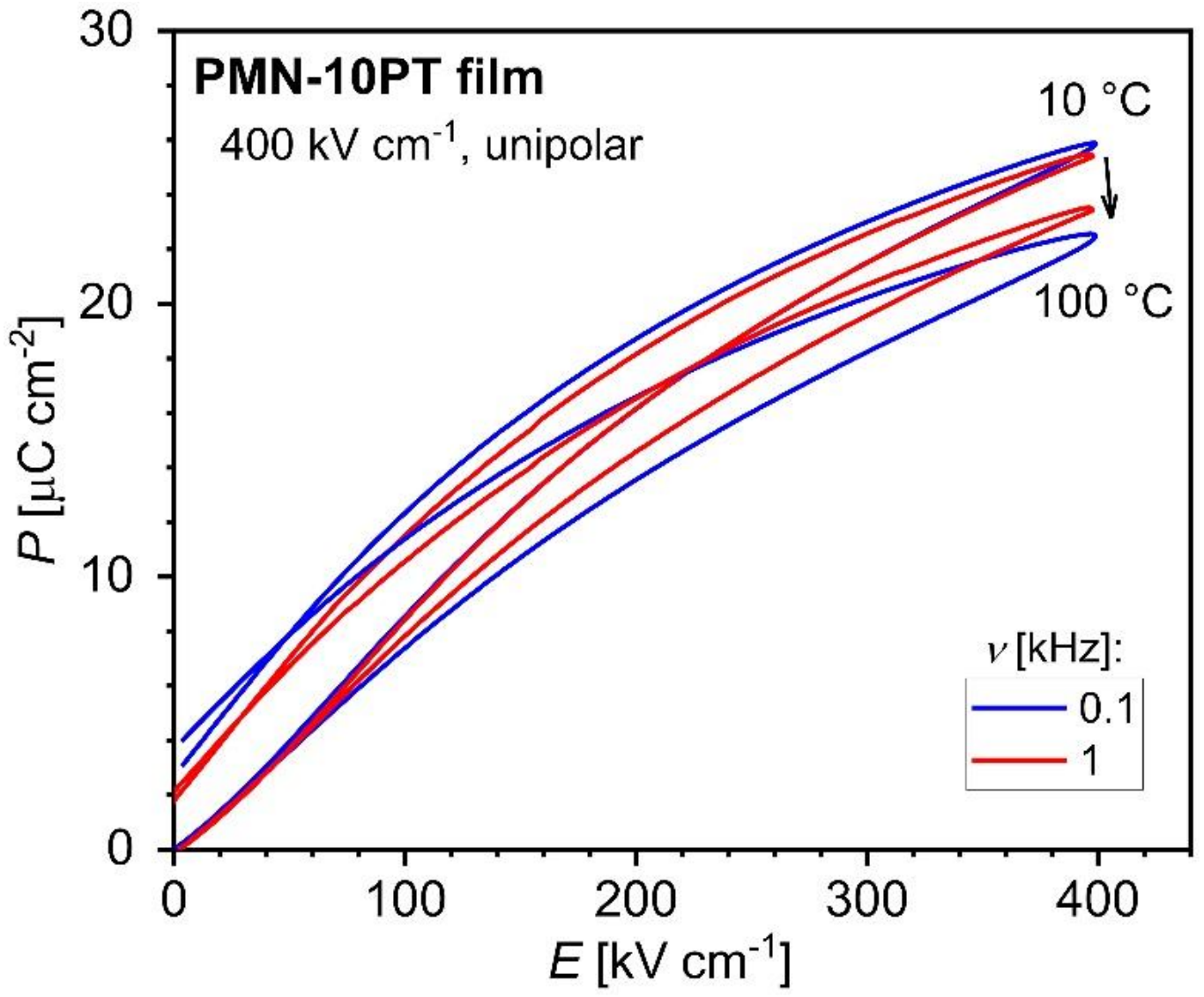

Figure S16. Unipolar *P*–*E* loops of PMN–10PT film measured at different frequencies and temperatures at 400 kV cm$^{-1}$.

In addition to the *P*(*T*,*E*) data, temperature-dependent $c_p$ data are needed to calculate the EC effect. The $c_p$ values of PMN–10PT were determined from bulk ceramics under conditions similar to those previously described for measurements on polyimide substrates (see SI, S6). Measurements were performed on disc-shaped plates (2*r* ~6 mm, *d* ~0.2 mm) in a temperature range of 10–110 °C at heating and cooling rates of 5 K min$^{-1}$. The $c_p(T)$ values shown in Fig. S17 were obtained from the second heating DSC run considering the known $c_p$ values of the standard sapphire material.[22] Since the PMN–10PT does not exhibit any phase transitions in the measured temperature range (i.e., 10–110 °C) and thus no singularities in the $c_p(T)$ data (Fig. S17), the electric-filed dependence can be neglected. Therefore, equation (4) can be simplified by placing the $c_p$ term in front of the integral.

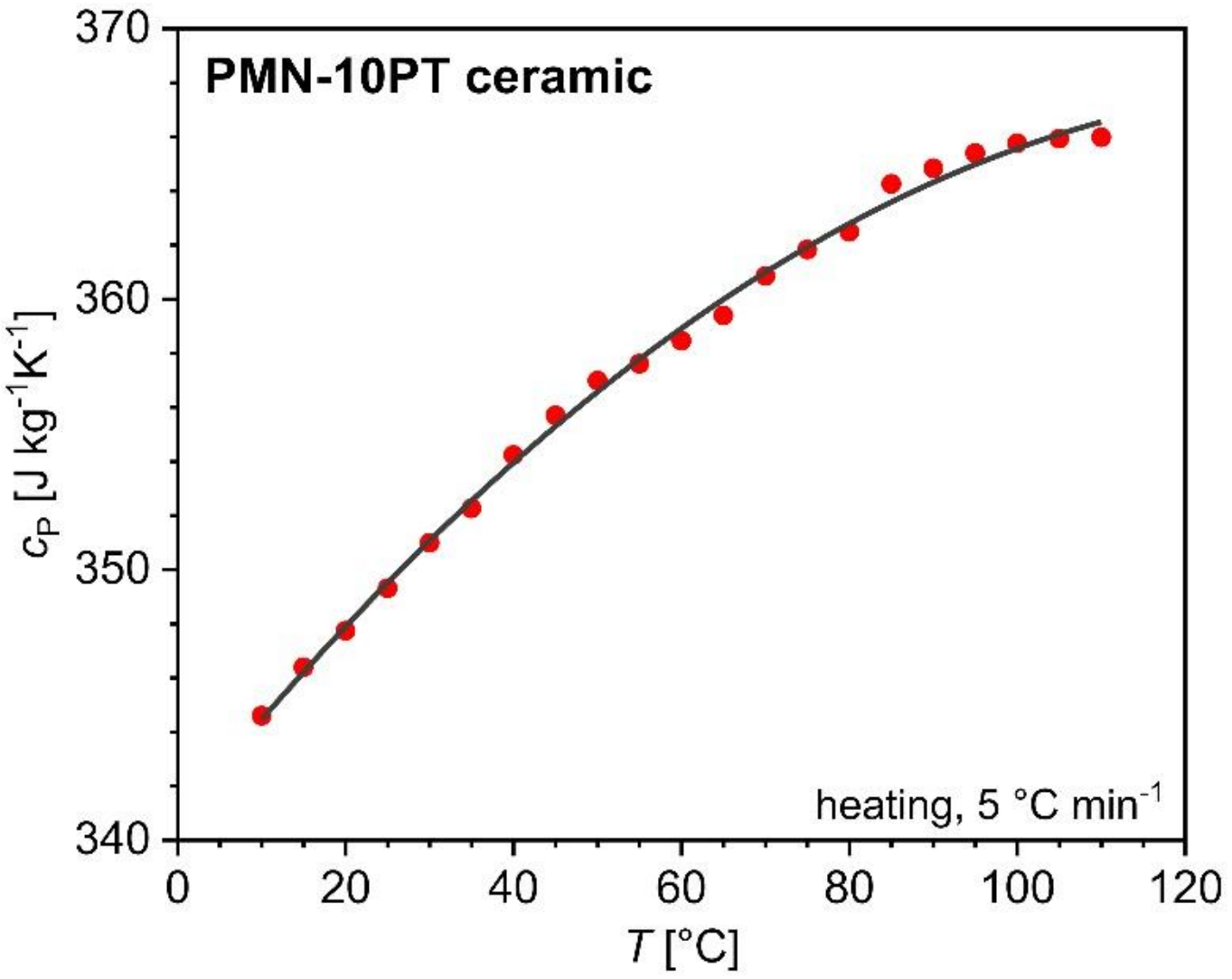

Figure S17. Temperature dependence of $c_p$ for PMN–10PT bulk ceramics.

Finally, by deriving and numerically integrating the $P(T,E)$ data and using $c_p(T)$ in equations (3) and (4), we can derive the $\Delta S_{EC}$ (Fig. S18a) and the $\Delta T_{EC}$ (Fig. S18b) of the PMN–10PT ceramic films, respectively. Both $\Delta S_{EC}$ and $\Delta T_{EC}$ (given here as absolute values) for EC cooling show a monotonic increase over the entire measured temperature range, leading to $\Delta S_{EC}$ of ~1.01 J $Kg^{-1}K^{-1}$ and $\Delta T_{EC}$ of ~0.86 K at 25 °C and 400 kV $cm^{-1}$, while their maximum of ~1.47 J $Kg^{-1}K^{-1}$ and ~1.54 K, respectively, is reached at 110 °C and 400 kV $cm^{-1}$.

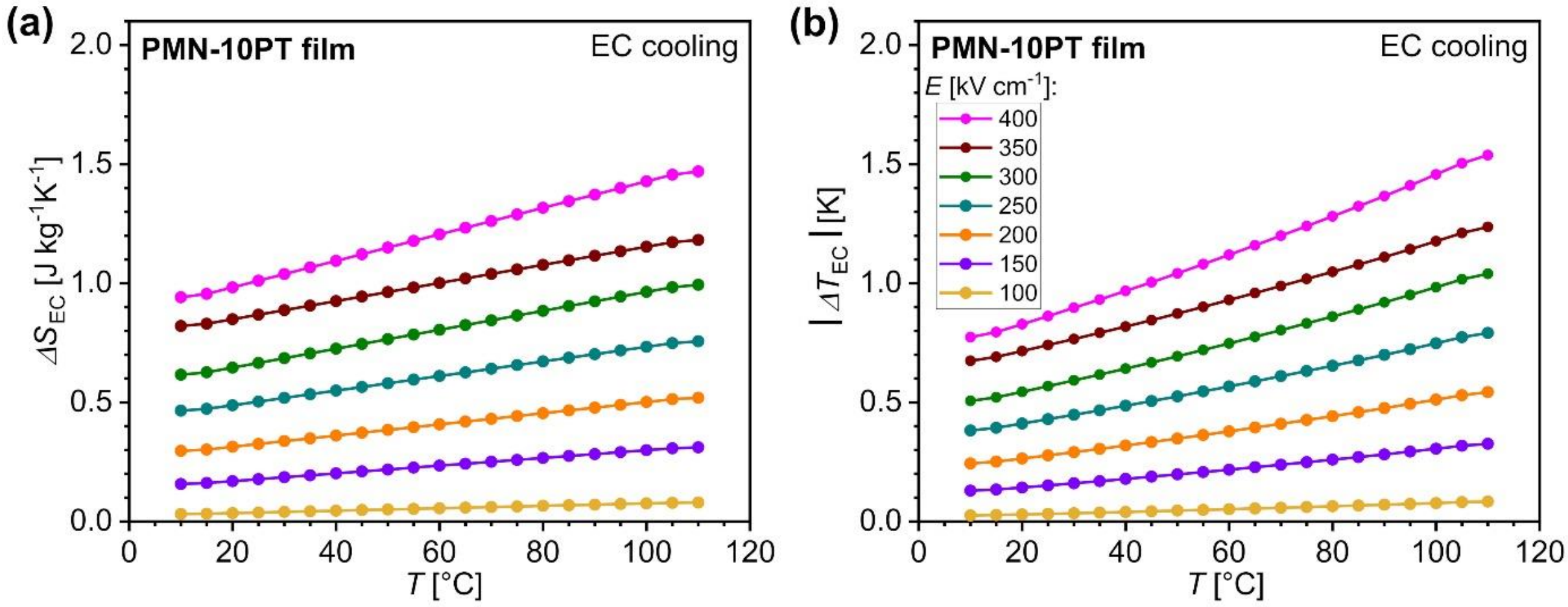

Figure S18. Temperature- and field-dependent (a) $\Delta S_{EC}$ and (b) $\Delta T_{EC}$ for electrocaloric cooling of the PMN–10PT film determined by the indirect Maxwell method.